\documentclass{ws-ijmpe}
\usepackage{url}
\usepackage[super,compress]{cite}
\usepackage{subfig}
\usepackage{float}

\newcommand{\ben}{\begin{displaymath}}
\newcommand{\een}{\end{displaymath}}
\newcommand{\be}{\begin{equation}}
\newcommand{\ee}{\end{equation}}
\newcommand{\bea}{\begin{eqnarray}}
\newcommand{\eea}{\end{eqnarray}}

\newcommand{\eq}[1]{Eq.~(\ref{#1})}

\newcommand{\xB}{\mbox{$x_p$}}
\newcommand{\etal}{{\it et al.{}}}

\begin{document}

\markboth{O. Hen et al.}{The EMC Effect and High Momentum Nucleons in Nuclei}

\catchline{}{}{}{}{}

\title{The EMC Effect and High Momentum Nucleons in Nuclei}

\author{\footnotesize Or Hen}
\address{Tel Aviv University, Tel Aviv 69978, Israel\\
or.chen@mail.huji.ac.il}

\author{\footnotesize Douglas W. Higinbotham}
\address{Thomas Jefferson National Accelerator Facility, Newport News, Virginia 23606, USA\\
doug@jlab.org}

\author{\footnotesize Gerald A. Miller}
\address{Department of Physics, University of Washington, Seattle, WA 98195-1560, USA\\
miller@phys.washington.edu}

\author{\footnotesize Eli Piasetzky}
\address{Tel Aviv University, Tel Aviv 69978, Israel\\
eip@tauphy.tau.ac.il}

\author{\footnotesize Lawrence B. Weinstein}
\address{Old Dominion University, Norfolk, Virginia 23529, USA\\
lweinste@odu.edu}

\maketitle

\begin{history}
\received{1/4/2013}
\end{history}

\begin{abstract}
  Recent developments in understanding the influence of the nucleus on
  deep-inelastic structure functions, the EMC effect, are reviewed. A
  new data base which expresses ratios of structure functions in terms
  of the Bjorken variable $x_A=AQ^2/(2M_A q_0)$ is presented.
  Information about two-nucleon short-range correlations from
  experiments is also discussed and the remarkable linear relation
  between short-range correlations and teh EMC effect is
  reviewed.  A convolution model that relates the underlying source of
  the EMC effect to modification of either the mean-field nucleons or
  the short-range correlated nucleons is presented.  It is shown that
  both approaches are equally successful in describing the current EMC
  data.
\end{abstract}


 


\section{Introduction}

Basic models of nuclear physics describe the nucleus as a collection of free nucleons moving non-relativistically under the influence of the sum of two-nucleon forces, which can be treated approximately as a mean field.
In this picture, in the rest frame of the nucleon, the partonic structure functions of bound and free nucleons should be identical. Therefore, it was generally  expected that, except for nucleon Fermi motion effects, 
Deep Inelastic Scattering (DIS) experiments which are sensitive to the partonic structure function of the nucleon would give the same result for all nuclei. 

Instead, the measurements show a reduction in the structure function
of nucleons bound in nuclei relative to nucleons bound in deuterium --
the EMC effect. Since its discovery, over 30 years ago, a large
experimental and theoretical effort has been put into understanding
the origin of the effect. While theorists have had no difficulty in
creating models that qualitatively reproduce nuclear DIS data by
itself, there is no generally accepted model. This is because the
models are either not consistent with or do not attempt to explain
other nuclear phenomena.  
The use of most modern models shows  that while
traditional nuclear effects such as binding and Fermi motion
contribute to the EMC effect, modification of the bound nucleon
structure is also required.

Studies of the effects of the many-body nucleon-nucleon interaction on
the structure of the nucleus predict the existence of Short-Range
Correlated (SRC) pairs.  These are pairs of nucleons at short distance
whose wave functions strongly overlap, giving them large relative
momentum and low center of mass momentum, where high and low is
relative to the Fermi momentum ($k_F$) of the nucleus.  Recent studies
show that the magnitude of the EMC effect in any nucleus is linearly
related to the number of two-nucleon SRC pairs in that nucleus.  The
observation of this phenomenological relationship raises a question of
whether the medium modification of the nucleon structure is related to
the nuclear mean field or to the SRC pairs. The answer to this
question will give new insight regarding the origin of the EMC effect.

Sections~\ref{sec:EMC} and~\ref{sec:SRC} review the EMC and SRC
research history respectively. Section~\ref{sec:EMC_Data} presents a
new formalism to correct the measured EMC data for the difference in
the definition of the Bjorken scaling variable for different nuclei
(the corrected EMC data is presented in~\ref{Appendix:EMC_DataBase}).
Section~\ref{sec:EMC-SRC} presents the EMC-SRC correlation, its
implications, and a simple convolution model which compares treatments
of the EMC effect based on nucleon modification occurring in SRC pairs
with that based on nucleon modification occurring due to the mean
field.
Section~\ref{sec:Summary} summarizes the paper. 



\section{The Nuclear EMC Effect}
\label{sec:EMC}

\subsection{Historical Overview}
\label{sec:EMC_History}

Unpolarized inclusive lepton scattering depends on two independent
variables that can be chosen as the negative of the square of the
transferred four-momentum, $Q^2=-q^2$ and the Bjorken scaling variable
for a proton $x_p=Q^2/(2 m_p \omega)$, -- commonly noted as $x_B$ -- 
where $m_p$ is the proton mass and $\omega$ the transferred energy in the proton rest frame.  
In Deep Inelastic Scattering (DIS), the
momentum transfer is large ($Q^2>2$ (GeV/c)$^2$) and the invariant mass of the
transferred photon plus the target nucleon is greater than the masses
of individual nucleon resonances, $W>2$ GeV. 
This allows a  measurement of the proton's
inelastic structure function, $F_{2}^{p}(x_p,Q^2)$, which gives the weighted average 
of the proton quark distribution function:
\bea
{  F_2^p(x_p,Q^2) = x_p \sum\limits_{q} e_q^2 \cdot  ( q^p(x_p)+\bar{q}^p(x_p) )    },
\label{eq:F2_def}\eea
where $q^p(x_p)$ and $\bar{q}^p(x_p)$ are the proton's quark and anti-quark distribution functions respectively, 
$e_q$ is the electric charge of the quark, and the sum runs over $q$ - the different proton quark flavours (i.e. $u$, $d$, and $s$).
The neutron inelastic structure function, $F_2^n(x_p,Q^2)$, is given by substituting in Eq.~\ref{eq:F2_def} the proton quark distributions by that of the neutron. 
The latter can be expressed using the proton distribution, assuming isospin (charge) symmetry (i.e. $u^n(x_p)=d^p(x_p)$, $d^n(x_p)=u^p(x_p)$ etc.).

\begin{figure}[tbph!]
\centering
\includegraphics[width=10cm, height=7.5cm]{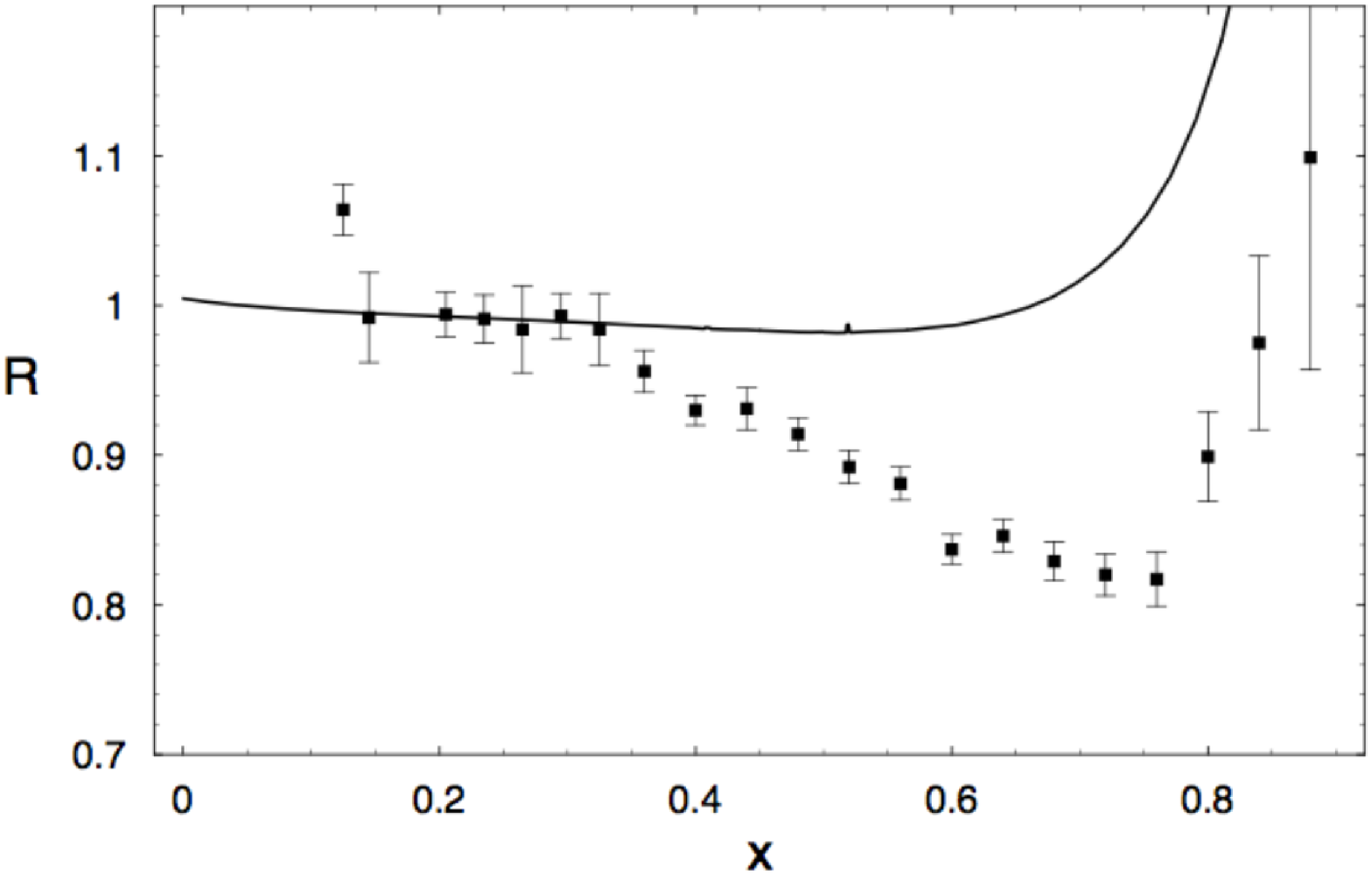}
\caption{Measurements of the DIS cross section ratio of gold relative to deuterium as a function of Bjorken-x$_p$ from SLAC. The solid black line is the expected ratio taking into account only Fermi motion of nucleons in Gold. 
(Figure reprinted from~\cite{Smith:2002ci}.  Copyright (2002) by the American Physical Society.)}
\label{fig:EMC}
\end{figure}

In the early 1980's, CERN's European Muon Collaboration (EMC) measured the per-nucleon DIS
cross section for scattering unpolarized muons from deuterium and iron
nuclei and extracted the ratio of their structure
functions~\cite{Aubert:1983xm} (i.e., $F_2^{Fe}(x_p,Q^2)/F_2^d(x_p,Q^2)$). 
The latter are the average $bound$ nucleon structure function in $^{56}Fe$ and Deuterium.
For $x_p\le0.5\sim0.7$, where nucleon Fermi motion effects
are negligible, they expected to measure a ratio of unity, indicating that the structure 
function of deeply bound (i.e. $^{56}Fe$) and loosely bound (i.e. Deuterium) nucleons is identical.  
This would allow them to increase the experimental luminosity by using a denser
target material such as iron, while still being sensitive to the free
nucleon structure function.  Instead, they discovered that the
per-nucleon DIS cross section ratio, which equals the structure function ratio, 
decreased from about 1 at $x_p\approx0.3$ to as little as 0.8 at $x_p\approx0.7$ (see
Fig.~\ref{fig:EMC}).  This unexpected result instantly became known as
the EMC effect. The existence of the EMC effect was soon verified by
analysis of existing target end-cap data from 
SLAC~\cite{Bodek:1983a,Bodek:1983b}, and later by
measurements at  SLAC~\cite{Arnold:1983mw} and the BCDMS and NMC
collaborations~\cite{Dasu:1988ru,Bari:1985ga,Amaudruz:1991cca}.

A later experiment performed at SLAC showed that the EMC effect has
the same qualitative behaviour for all nuclei, differing only in the
value of the ratio at the minimum~\cite{Gomez:1993ri}.  It also showed
that the EMC effect is independent of $Q^2$ for $2\le Q^2\le40$
(GeV/c)$^2$ and that the depth of the minimum at $x_p\approx0.7$ grows
with nuclear mass.  The growth seemed to increase with the average
nuclear density~\cite{Gomez:1993ri} and this became a generally
accepted feature of the EMC effect (see Ref. ~\cite{Geesaman:1995yd}
and references therein).

As theorists provided several different,  simple, explanations of the
effect (see discussion in section~\ref{sec:EMC_Theory}), an
independent experimental test of these explanations was needed~\cite{Bickerstaff:1985ax}. 
This came from Fermi National Accelerator Lab in the form of Drell-Yan measurements~\cite{Alde:1990im}. These experiments compared $\mu^+-\mu^-$ production from $q-\bar q$ annihilation in proton-proton and proton-nucleus collisions.
In the kinematic range covered by the measurement, they observed that the nuclear to proton ratio  was consistent with unity. 
As this experiment was sensitive mainly  to the nuclear sea quarks, the result pointed to the EMC effect being due to a change in the valence quark distributions.

\begin{figure}[tbph!]
\centering
\includegraphics[width=8.5cm, height=6.5cm]{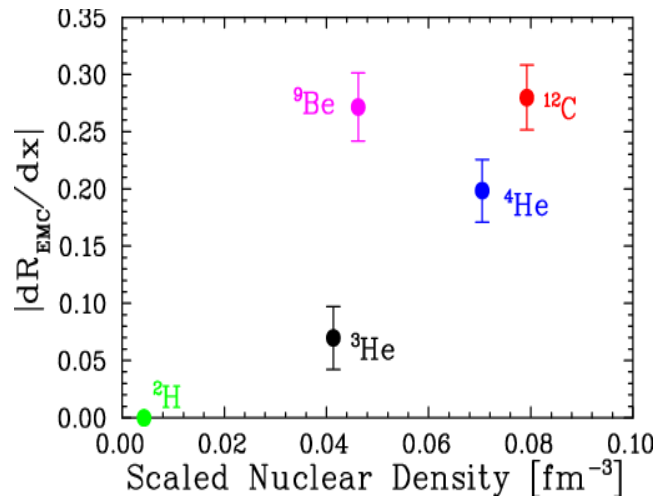}
\caption{The strength of the EMC effect, defined as the slope of the per nucleon DIS cross section ratio for $0.3\le x\le0.7$, shown as a function of the scaled nuclear density for light nuclei. 
(Figure reprinted from~\cite{Seely:2009gt}.  Copyright (2009) by the American Physical Society.)}
\label{fig:SeelySlopes}
\end{figure}

It was not until 2009 that the simple nuclear density dependence of
the EMC effect was challenged with new data~\cite{Seely:2009gt}.  A
high precision measurement on light nuclei, including
$^3$He, $^4$He, $^9$Be and $^{12}$C, showed that the effect was not
related to the average nuclear density.  The most significant outlier
was $^9$Be which has a low average nuclear density, similar to that of
$^3$He, and a large EMC effect, similar to that of $^4$He and $^{12}$C
(see Fig.~\ref{fig:SeelySlopes}).  This anomaly was consistent with
variational Monte Carlo calculations which show that local, high
density configurations occur in nuclei~\cite{Pieper:2001mp}. These
calculations describe $^9$Be as a collection of two alpha clusters and
an orbiting neutron.  In this picture, $^9$Be has a low average
density and a much higher local density similar to that of $^4$He. Thus, the
phenomenological explanation of the EMC effect shifted, based on the
new data, from an average density effect to a local density effect.

\subsection{Theory Status}
\label{sec:EMC_Theory}

In QCD, the nucleon structure function, $F_2(x,Q^2)$, gives the
weighted probability for finding a parton (quark) in the nucleon that
carries a fraction $x$ of the total nucleon momentum. The different
partons contribute with a weight equal to the square of their electric charge.
The primary  theory interpretation of the reduction of the nuclear structure function in the valence quark region was simple~\cite{Frankfurt:1988nt,Arneodo:1992wf,Geesaman:1995yd,Piller:1999wx,Sargsian:2002wc,Miller:2007zzb}.
Quarks in nuclei carry less  momentum than quarks in nucleons  and, as the uncertainty principle implies,  move  throughout  a larger confinement volume. This notion gave  rise to a host of models: bound nucleons are larger than free ones; 
quarks in nuclei move in 6 quark or 9 quark or even $3A$ quark bags~\cite{Jaffe:1982rr,Carlson:1983fs,Pirner:1980eu}. But more conventional explanations such as the influence of nuclear binding or enhancement of pion cloud effects were successful in reproducing some of the nuclear deep inelastic scattering data~\cite{Ericson:1983um,Berger:1987er,Llewellyn Smith:1983qa,Jung:1988jw,Jung:1990pu}. And one could combine various different models~\cite{Bickerstaff:1985ax,Bickerstaff:1985da}. This led to  a plethora of models that reproduced
the data, causing one of the present authors to write that EMC means Everyone's Model is Cool\cite{Miller:1988hj}. 
It is interesting to note that none of the earliest models were   concerned with the role of  two nucleon correlations, except as relating to  six quark bags.

The initial excitement tapered off as nuclear deep inelastic scattering became more understood, the experimental data became more precise, and the need to include  the effects of nuclear shadowing was acknowledged~\cite{Frankfurt:1988zg}.  Indeed some of the more extreme models were ruled out by a failure to match well-known nuclear phenomenology. Moreover, inconsistency with  the baryon and momentum sum rules led to the  downfall of many models~\cite{Frankfurt:1985ui}.    Some models predicted an enhanced nuclear sea, but others did not. As results from Drell-Yan measurements were published, none of the existing models survived the challenge of providing an accurate description of both the EMC and Drell-Yan data sets --  a challenge that remains to this day. 
   
It is now understood that conventional nuclear binding effects can account for the EMC effect up to values of
$x_p\approx0.5$ or so~\cite{CiofidegliAtti:1991ae,Dieperink:1991mw,Frankfurt:2012qs,Miller:2001tg,Smith:2002ci} but fail at larger values.  Therefore the effects of the nuclear modification of
the nucleon structure function must be included. Currently viable
models of nucleon modification include (a) the quark meson coupling model in which quarks in nucleons (either bags or eigenstates of the NJL model)~\cite{Mineo:2003vc,Cloet:2006bq} exchange mesons  with 
quarks in other nucleons, (b) the chiral quark soliton model  in which
quarks in nucleons also exchange mesons with other nucleons~\cite{Smith:2003hu}, and (c) the suppression of point-like-configurations of the nucleon by the nuclear medium~\cite{Frankfurt:1985cv,Frank:1995pv}.
A successful phenomenology that includes the effects of shadowing, binding, pion enhancement and a medium-modification of the quark structure function  can be fitted to the extant data~\cite{Kulagin:2004ie}.

A modern model which incorporates the influence of nucleon-nucleon
correlations in a manner consistent with nuclear physics knowledge to
describe both nuclear deep inelastic scattering and Drell-Yan data   does not yet exist.

\subsection{EMC Data Analysis}
\label{sec:EMC_Data}
Following the EMC collaboration, other experiments measured the ratios
of per-nucleon DIS cross sections for nuclei and the deuteron at equal values of
$Q^2$ and $x_p = Q^2/2m_p\omega$. In these kinematical conditions, 
the DIS cross section ratio for nuclei $A_1$ and $A_2$ is given by~\cite{Halzen:1984mc}:
\bea
{  \frac{\sigma_{A_1}}{\sigma_{A_2}}=\frac{F_2^{A_1}(x_p,Q^2)}{F_2^{A_2}(x_p,Q^2)} \cdot \frac{\left[ 1+2\frac{1+\omega^2/Q^2}{R_{A_1}-1}\text{tan}^2\frac{\theta}{2} \right]}{\left[ 1+2\frac{1+\omega^2/Q^2}{R_{A_2}-1}\text{tan}^2\frac{\theta}{2} \right]} },
\label{eq:DIS_xSection}\eea
where $\theta$ is the lepton scattering angle and $R_A=\sigma_L^A/\sigma_T^A$ is the ratio of the longitudinal to transverse cross section for nucleus A.
Assuming 
$R_A$ independent of $A$, the cross section ratio of Eq.~\ref{eq:DIS_xSection} is reduced to the the $F_2$ structure function
ratio.

Recently, Frankfurt $\&$ Strikman (FS) pointed out that the structure functions of nucleons bound in nuclei should be extracted in the reference frame of the nucleus~\cite{Frankfurt:2012qs}. This is done by using the $x_A$ scaling variable, defined as:
\bea
{   x_A=\frac{Q^2}{2 q \cdot P_A / A}=\frac{AQ^2}{2\omega m_A}=x_p \cdot \frac{Am_p}{m_A}    },
\label{eq:xA}\eea
where $q$ and $P_A$ are the 4-momentum vectors of the virtual photon
and target nucleus respectively, and $m_A$ is the mass of the target
nucleus.  Note that, for the same values of $Q^2$ and $\omega$, $x_A$
differs from $x_p$ by the ratio of the bound nucleon mass to the free
mass.  Therefore, a cross section measured at $Q^2$ and $\omega$ on
nucleus $A$ will depend on the nucleon structure function evaluated at
$x_A$ rather than at $x_p$.  

This means that the standard EMC cross section ratio at the same $Q^2$
and $\omega$ (and hence the same $x_p$) is actually proportional to
the nucleon structure function in nucleus $A$ evaluated at parton
momentum fraction $x_A=AQ^2/2m_A\omega$ divided by the nucleon structure function in
deuterium evaluated at parton momentum fraction $x_d=2Q^2/2m_d\omega$.  For symmetric
nuclei this is: 
\bea { \frac{2}{A}
  \frac{\sigma_{DIS}^A(x_p,Q^2)}{\sigma_{DIS}^d(x_p,Q^2)} =
  \frac{F_2^A(x_A,Q^2)}{F_2^d(x_d,Q^2)} }
\label{eq:EMC_xA}
\eea
where $\frac{\sigma_{DIS}^A}{\sigma_{DIS}^d}$ is the DIS cross section
ratio measured at the same $(Q^2,\omega,x_p)$, and
$\frac{F_2^A(x_A,Q^2)}{F_2^d(x_d,Q^2)}$ is the ratio of structure
functions at the same $(Q^2,\omega)$ but different $x$.  Since we want
to compare the structure functions at the same parton momentum
fractions, we want to correct this using 
\bea {
  \frac{F_2^A(x_A,Q^2)}{F_2^d(x_d,Q^2)}=\frac{F_2^A(x_A,Q^2)}{F_2^d(x_A,Q^2)}
  \cdot \frac{F_2^d(x_A,Q^2)}{F_2^d(x_d,Q^2)} },
\label{eq:EMC_xA}
\eea
and
\bea
\frac{F_2^A(x_A,Q^2)}{F_2^d(x_A,Q^2)} = {  \frac{\sigma_{DIS}^A(x_p,Q^2)}{\sigma_{DIS}^d(x_p,Q^2)} \cdot \frac{F_2^d(x_d,Q^2)}{F_2^d(x_A,Q^2)}   },
\label{eq:EMC_F2ratio}
\eea
where
$\frac{F_2^A(x_A,Q^2)}{F_2^d(x_A,Q^2)}$ is the ratio of structure
functions in the different nuclei evaluated at the same parton
momentum fraction (i.e., the quantity we wish to extract), and $\frac{F_2^d(x_A,Q^2)}{F_2^d(x_d,Q^2)}$ is a correction factor. This correction factor can be evaluated using well known parameterizations of the deuteron structure function~\cite{Whitlow:1992,Bosted:2008}.

\begin{figure}[tbph!]
\centering
\includegraphics[width=9cm, height=7cm]{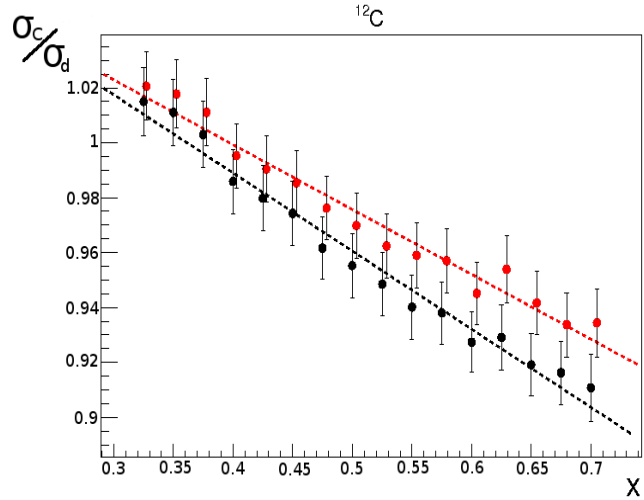}
\caption{Example of the effect of the $x_A$ correction to the
  data. The black points are the original Seely data plotted as a
  function of $x_p$. The red points are the corrected data, plotted as
  a function of $x_A$. Dashed lines are linear fits to the two data
  sets. The difference between the slope of the two fits is about $20\%$.}
\label{fig:x correction}
\end{figure}

Fig.~\ref{fig:x correction} shows the effect of the correction factor of Eq.~\ref{eq:EMC_F2ratio} on the measured DIS cross section ratio for $^{12}$C relative to deuterium from recent Jefferson Lab measurements.
As can be seen, the $x_A$ correction reduces the size of the EMC
effect (i.e., its slope).  It replaces part of the model-dependent
binding energy corrections with a systematic, transparent, and
model-independent correction.

For asymmetric nuclei ($N\neq Z$), following Aubert {\it et al.} and
Bodek {\it et al.}~\cite{Aubert:1983xm,Bodek:1983a,Bodek:1983b}, an additional isoscalar correction factor ($R_{ISO}$) is applied to the measured cross section ratio, making it related to a hypothetical nucleus with equal number of protons and neutrons ($N=Z=A/2$):
\bea 
{  \frac{\sigma_{DIS}^A(x_A,Q^2)_{ISO}}{\sigma_{DIS}^d(x_A,Q^2)_{ISO}} =  \frac{\sigma_{DIS}^A(x_A,Q^2)}{\sigma_{DIS}^d(x_A,Q^2)} \cdot R_{ISO}(x_A)},
\label{eq:EMC_xB_ISO}\eea
with $R_{ISO}(x_A)$ defined as:
\bea 
{   R_{ISO}(x_A)=\frac{A}{2}  \frac{F_2^p(x_A,Q^2)+F_2^n(x_A,Q^2)}{Z \cdot F_2^p(x_A,Q^2)+N \cdot F_2^n(x_A,Q^2)}=\frac{A}{2}  \frac{1+R_{np}(x_A,Q^2)}{Z+N \cdot R_{np}(x_A,Q^2)}  },
\label{eq:EMC_ISO}\eea
where $F_2^p(x_A,Q^2)$ and $F_2^n(x_A,Q^2)$ are the free proton and neutron structure functions, and $R_{np}(x_A,Q^2)={F_2^n(x_A,Q^2)}/{F_2^p(x_A,Q^2)}$. The free neutron structure function used in this correction is usually extracted from world data on DIS scattering off deuterium and the proton, corrected for the Fermi motion of protons and neutrons in deuterium (i.e., smearing effect), see~\cite{Bodek:1979,Arrington:2009} for details. 

Using the isoscalar correction for asymmetric nuclei and the description of the measured cross section ratios in terms of $F^A_2$ and $F^d_2$ (Eq.~\ref{eq:EMC_xA}), we extract the structure function ratio of nucleons bound in nuclei relative to deuterium as:
\bea
  \frac{F_2^A(x_A,Q^2)}{F_2^d(x_A,Q^2)}&=&\frac{2}{A} \frac{\sigma_{DIS}^A(x_p,Q^2)}{\sigma_{DIS}^d(x_p,Q^2)} \cdot \frac{F_2^d(x_d,Q^2)}{F_2^d(x_A,Q^2)} \cdot  R_{ISO}(x_A)  =     \nonumber \\ 
&=&\frac{2}{A}   \frac{\sigma_{DIS}^A(x_p,Q^2)_{ISO}}{\sigma_{DIS}^d(x_p,Q^2)_{ISO}} \cdot   \frac{F_2^d(x_d,Q^2)}{F_2^d(x_A,Q^2)}  \cdot   \frac{ R_{ISO}(x_A)}{R_{ISO}(x_p)}    ,
\label{eq:EMC_xACorrection}\eea

~\ref{Appendix:EMC_DataBase} presents the EMC ratios, extracted as a function of $x_A$ for $0.3 \lesssim x_A \lesssim 0.7$, for all nuclei measured at SLAC~\cite{Seely:2009gt} and JLab~\cite{Gomez:1993ri}. The isoscalar correction applied is identical for both data sets, making them more consistent.

\section{High momentum nucleons in nuclei}
\label{sec:SRC}
This section describes nucleon-nucleon (or ``two-nucleon'')
correlations. A correlated two-nucleon pair is one where the
two-nucleon density is significantly different from the
product of two single-nucleon densities. Both tensor and central
forces can produce short-range correlations.

\subsection{Theoretical Need for High Momentum Nucleons in Nuclei}
\label{sec:SRC_Theory}

The strong interactions between nucleons in nuclei are dominated by
two and three nucleon terms. Therefore the fact that nucleons in
nuclei are correlated is self-evident. There is no fundamental
one-body potential in the nucleus, unlike the one-body Coulomb
potential in atomic physics. The fundamental question of nuclear physics was: how
does the very successful shell model of the nucleus emerge in spite of
the strong short-ranged interactions between nucleons? An answer was
provided early on by Brueckner \& Goldstone, see the review by
Bethe~\cite{Bethe:1971xm}. The strong two-nucleon interactions encoded
by the potential $V$, constructed to reproduce experimentally measured
scattering observables and believed to include strong repulsion at
short distance and attraction at longer ranges, are summed to form the
$T$ matrix of scattering theory and the $G$-matrix for bound
states. The operator $G$ is obtained from $T$ by modifying the
propagator to include the effects of the Pauli principle and to use
the appropriate self-consistent (single) nucleon energies. The $G$
matrix is considerably weaker than $V$, and can therefore be used in
perturbation theory. One forms the nuclear mean field $U$ throughout
the Hartree-Fock method employing the $G$-matrix, and the first
approximation to the wave function is the anti-symmetrized product of
single particle wave functions engendered by $U$. However, the
complete nuclear wave function is obtained in a perturbative hole-line
expansion that includes two-particle -- two-hole excitations and other
excitations which incorporate correlations. Later work formulated a
relativistic version of Brueckner theory in which the Dirac equation
replaces the Schroedinger
equation~\cite{Anastasio:1984gy,Brockmann:1984qg}. There is also a
light front version~\cite{Miller:1999ap,Miller:2000kv}.

The Brueckner theory approach described above presumes that the
two-nucleon potential contains strong short-distance repulsion.  Early
attempts to construct soft potentials lacking the strong repulsion
that also reproduce scattering data did not succeed in obtaining
interactions that could be used perturbatively in the nuclear bound
state problem~\cite{Bethe:1971xm}. In modern times, the use of
effective field theory provides a low-energy version of QCD, guided by
chiral symmetry, in which one obtains the potential as an expansion in
powers of $(Q/\Lambda_\chi)$ where $Q$ is a generic external momentum
(nucleon momentum) and $\Lambda_\chi $ is the chiral symmetry breaking
scale of about 1 GeV. See the review~\cite{Bedaque:2002mn}. In such
theories the short distance interaction can be treated as a contact
interaction, modified by the inclusion of a cut-off, and the longer
ranged interactions are accounted for by one and two pion exchange
interactions. The softness (involving low momentum) or hardness
(involving higher momentum) of the potential is determined by the
value of the cutoff. For sufficiently soft potentials nuclear matter
can be treated using perturbation theory in terms of the two and three
nucleon chiral interactions. Nevertheless, two-nucleon correlations
occur, primarily as a result of the second iteration of the one pion
exchange potential.

Another approach uses renormalization group methods to generate a
soft $NN$ potential from a hard interaction either by integrating out
high momentum components (in the case of $V_{low-K}$), or by using the
similarity renormalization group~\cite{Bogner:2009bt}. Then one
obtains a potential that is mainly restricted to small values of
momentum. This potential is perturbative in the sense that the Born
series for scattering converges and perturbation theory can be applied
to the nuclear bound state problem. However, once again the
second-order term in the potential generates correlations.

The renormalization group can be used to eliminate matrix elements of the nucleon-nucleon potential
 connecting low and high relative momentum states. Such a procedure simplifies the computations of nuclear binding energies and spectra, and would also lead to wave functions without high-momentum components  and  truly short ranged-correlations.  However, it would be necessary to  consistently transform all other operators. For high momentum transfer reactions, the renormalization group changes a known simple probe, described by a single-nucleon operator into a complicated probe describable by unknown (in practice) $A$-nucleon operators. This prevents the analysis of any high momentum transfer experiment.
  
To summarize, there are two basic approaches to fundamental nuclear structure -- perturbation in the $G$ matrix or perturbation in the potential.
In either case there will be two-nucleon correlations.  Theoretically,
the key remaining question concerns the quantity and range of the
correlations.

\subsection{SRC Measurements}
\label{sec:SRC_Exp}

Experimentalists at electron scattering facilities such as SACLAY and
NIKHEF observed the need for high momentum components in nuclei, not
from direct observation, but rather from a dramatic lack of cross section
in $A(e,e'p)A$$-$$1$ valence shell knock-out experiments where the
independent particle models overestimated the measured cross
sections~\cite{Kelly:1996hd}.  Since the shell model accurately
predicts energy levels and spins, this reduced the range of possible
explanations.  The most straightforward explanation was that
the ``missing nucleons'' were in nucleon-nucleon correlations.  When
the electron scattered from a nucleon in a correlated pair, its
partner was also ejected from the nucleus.  This shifted strength from
excitation energies typical of valence states to much higher
excitation energies.

 Many experiments were done at these facilities to probe
for more direct evidence of correlations; but as history would show,
the necessary kinematic requirements, $x_B > 1$ and $Q^2 >
1$(GeV/c)$^2$, were practically inaccessible.  Thus most of the early
experiments ended up being studies of reaction mechanisms such as
meson-exchange currents and final-state interactions.

With the availability of continuous, high intensity, high momentum
proton and electron beams, identifying SRCs in Quasi-Elastic (QE)
scattering off nuclei became feasible.  In this section we review
results from measurements of inclusive QE $(e,e')$ cross section
ratios and exclusive, triple-coincidence, $(e,e'pN)$ and $(p,2pn)$,
large momentum transfer (hard), reactions performed at the Stanford
Linear Accelerator Center (SLAC), Brookhaven National Lab (BNL) and more
recently at Jefferson Lab (JLab).

\subsubsection{Inclusive SRC Measurements}
In inclusive scattering of unpolarized particles from an unpolarized
target, there are only two independent kinematical variables. In the
case of inclusive QE electron scattering these are normally chosen to
be $Q^2$ and $x_p$.  In the Plain Wave Impulse Approximation (PWIA) it
is assumed that the virtual photon is fully absorbed on a single
nucleon, which leaves the nucleus without rescattering, leaving the
remaining $A-1$ nuclear system unperturbed.
Energy and momentum conservation for such a reaction define a minimum
value for the component of the initial momentum of the scattered
nucleon in the direction of the virtual photon as a function of $Q^2$
and $x_p$~\cite{Egiyan:2003vg}.  At $x_p=1$, for all $Q^2$ values, the
minimum value of this momentum component equals zero. As one increases
or decreases $x_p$ at fixed $Q^2$, its value increases.  At
moderate values of $Q^2$ ($\sim2-4$ GeV/c$^2$) and $x_p\ge1.4-1.5$
($\le\sim0.6$) this minimum value is larger than the Fermi momentum
($k_F$) of the nucleus, and the reaction is dominated by scattering
from high momentum ($\ge k_F$) nucleons in the nucleus. At these $Q^2$
values and $x_p < 1$, the virtual photon carries a large amount of
energy compared to its momentum and the reaction, while sensitive to
high momentum nucleons, has large inelastic contributions from $\Delta$
production and meson exchange currents (MEC). On the
other hand, for the same $Q^2$ values and $x_p>1$, the virtual photon
transfers a small amount of energy compared to its momentum, inelastic
processes are suppressed, and the reaction is more directly sensitive
to the nature of the high momentum tail of the nuclear wave
function~\cite{Frankfurt:1997, Sargsian:2001}. In both cases, large
values of $Q^2$ suppress meson exchange current (MEC)
contributions~\cite{Arnold:1990, Laget:1987}.

\begin{figure}[tbp]
\centering
\includegraphics[width=8.5cm, height=9.5cm]{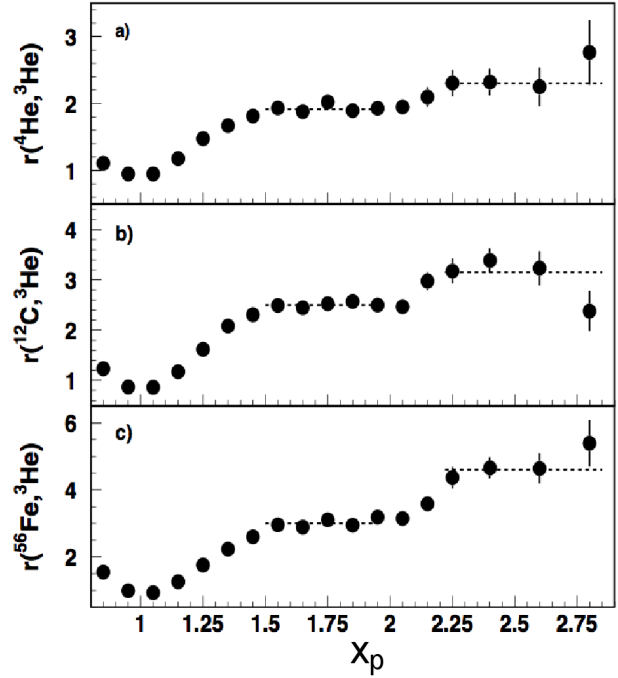}
\caption{Per nucleon QE inclusive $(e,e')$ scattering cross section ratios for nuclei relative to $^3$He plotted 
as a function of $x_p$. Two plateaus are observed for $1.5\le x_p\le2$ and $x_p\ge2.25$. The magnitude 
of these plateaus are labeled as $a_2$ and $a_3$, respectively. In the SRC model of the high momentum 
tail of the nuclear wave function, they are taken as a measure of the relative amount of 2N 
and 3N SRC pairs in the measured nuclei. See text for more details.
(Figure reprinted from~\cite{Egiyan:2006}.  Copyright (2006) by the American Physical Society.)}
\label{fig:Egiyan_a2}
\end{figure}

Inclusive electron scattering cross section ratios for nucleus $A$
relative to deuterium and to $^3$He were measured at SLAC and later at
Hall-B and Hall-C of JLab~\cite{Frankfurt:1993sp, Egiyan:2003vg,
  Egiyan:2006, Fomin:2012}.  Fig.~\ref{fig:Egiyan_a2} shows the $x_p$
dependence of the per nucleon cross section ratio of nuclei relative
to $^3$He measured at Hall-B.  As can be seen, for $x_p$ values which
correspond to scattering off high momentum ($\ge k_F$) nucleons in the
nucleus (i.e., $1.5\le x_p\le2$ and $x_p\ge2.25$) the cross section
ratio scales (i.e., does not depend on $x_p$). The contribution of
Final State Interactions (FSI) to the measured cross sections are
expected to rapidly decrease as a function of $Q^2$. Calculations of
FSI in inclusive scattering at large $Q^2$ and $x_p\ge1$ 
show they are largely confined to within the nucleons of the
initial-state SRC pair~\cite{Frankfurt:1997}.  The contribution of FSI
of this kind will cancel in the cross section ratio of two nuclei.
This is supported by the small observed $Q^2$ dependence of the cross
section scaling plateau.  This scaling reflects the scaling of the
high momentum tail of the nuclear wave function and is usually
interpreted using the Short-Range-Correlation (SRC)
model~\cite{Frankfurt:1993sp, CiofidegliAtti:1995qe}.  The latter
states that the high momentum tail of the nuclear wave function is
dominated by correlated, multi-nucleon, configurations. Due to their
strong interaction at short distances, the structure of these
configurations is independent of the surrounding nuclear environment,
resulting in the same shape of the high momentum tail in all nuclei
(i.e., scaling). Different nuclei have different 
amounts of short range correlated (SRC) clusters. In this model, the
observed scaling of the per-nucleon cross section ratios for $1.5\le
x_p\le2$ and $x_p\ge2.25$ are indicative of scattering off two-nucleon
(2N) and three-nucleon (3N) SRC configurations, respectively. The
scaling factors, noted as $a_2$ and $a_3$ are then a measure of the
relative amount of 2N and 3N SRC, respectively, in the measured
nuclei.

\subsubsection{Exclusive (p,2pn) and (e,e'pN) Measurements}
Inclusive measurements alone do not
prove that high momentum nucleons are a result of initial-state
short-range correlated pairs.  To study the contribution of 2N-SRC
pairs to the high momentum tail of the nuclear wave function exclusive
two-nucleon-knockout experiments were done. The concept behind such
experiments is that, in the Plain-Wave-Impulse-Approximation (PWIA), in the absence of FSI,
when a nucleon that is part of a 2N-SRC pair is knocked out the
nucleus, in order to conserve momentum its correlated partner nucleon
has to recoil with momentum that is equal in size and opposite in
direction to the initial momentum of the knocked-out nucleon. This
back-to-back correlation between the initial momentum of the
knocked-out nucleon and the momentum of the recoil nucleon, both above
the Fermi sea level $k_F$, is a clear signature for scattering off a
2N-SRC configuration.  Due to the center-of-mass (c.m.) motion of the
2N-SRC pair with respect to the residual $A-2$ nuclear system, this
correlation will not be exactly back-to-back.  The measured angular
correlation can be used to extract the c.m. momentum distribution of
the pair. If the 2N-SRC model is correct, the nucleons in the pair
will have large relative momentum ($\ge k_F$) and small c.m. momentum
($\le k_F$).

\begin{figure}[tbp]
  \centering
  \subfloat       {\label{fig:EC_tot_p:All} \epsfig{height=7cm,width=7cm,file=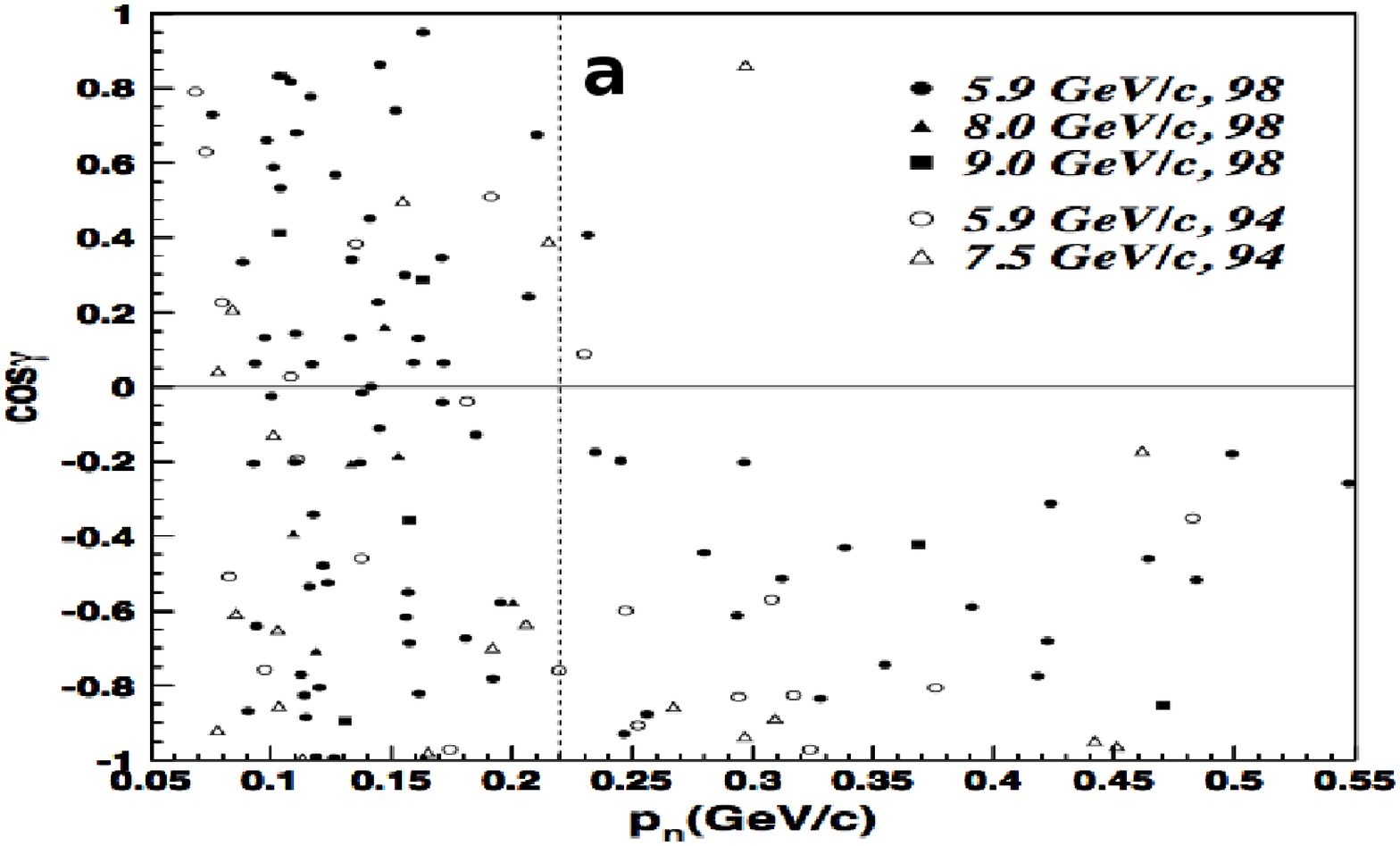}  \epsfig{height=6.5cm,width=4.5cm,file=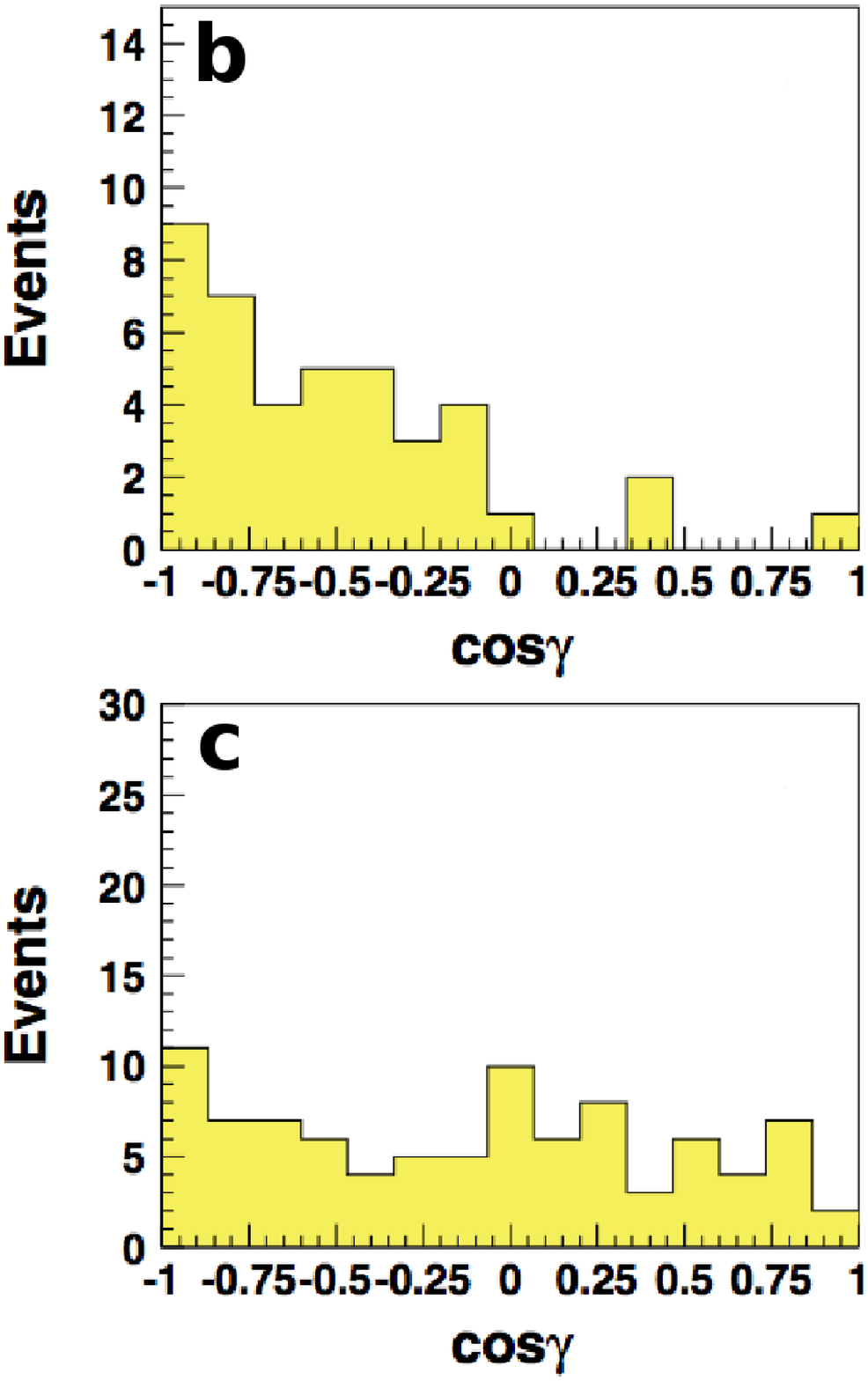}} \\
  \subfloat {\label{fig:EC_tot_p:Xcut} \epsfig{height=5.5cm,width=8.5cm,file=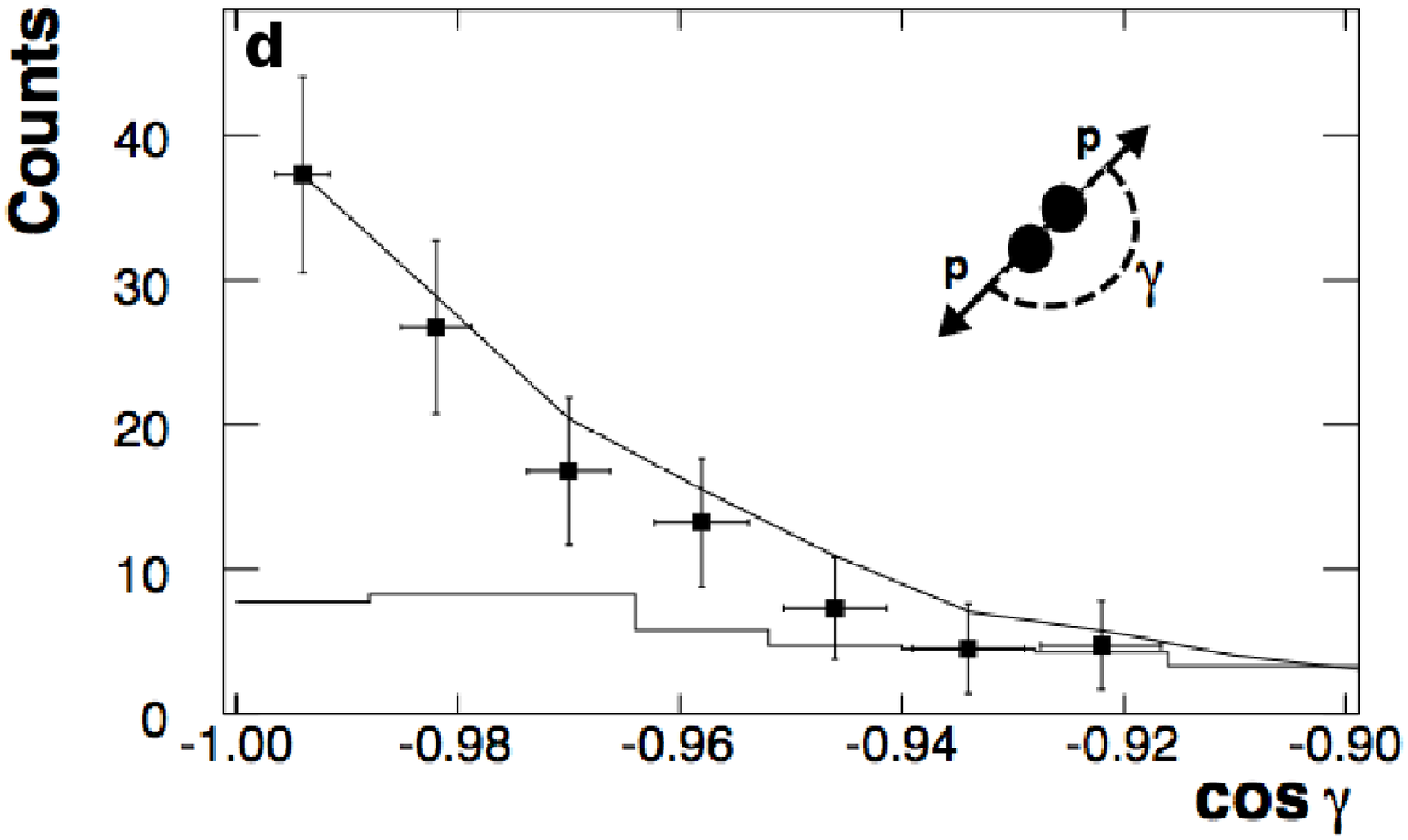}} \\ 
  \caption {Distributions of the relative angle ($\gamma$) between the reconstructed initial momentum of the knockout proton 
and the recoil nucleon. Top: Results for proton induced proton-neutron pair knockout (i.e. $^{12}$C$(p,2pn)$) measurements from BNL, shown and a function of: (a) the momentum 
of the recoil neutron, (b) for events with recoiling neutron with momentum greater then the Fermi momentum, and (c) for events with recoiling neutron with momentum lower then the Fermi momentum.  
These results show a clear transition from an isotropic distribution to a back-to-back correlated distribution as the recoil neutron momentum reaches the Fermi momentum of $^{12}$C.
Bottom: Results for electron induced proton-proton pair knockout (i.e. $^{12}$C$(e,e'pp)$) measurements from JLab, shown for events in which the initial momentum of the knockout proton, $|\vec{p}-\vec{q}|$, equals $\sim500$ MeV/c.
(Figures reprinted from~\cite{Tang:2003,Piasetzky:2006ai,Shneor:2007}.  Copyright (2003, 2006, 2007) by the American Physical Society.)}
\label{fig:OpeningAngle}
\end{figure}

\begin{figure}[tbp]
  \centering
    \includegraphics[width=14cm, height=6.5cm]{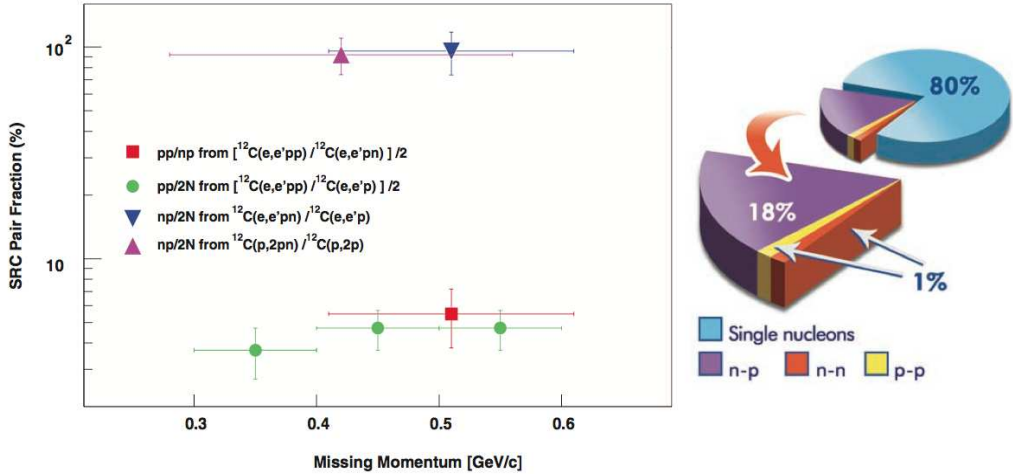}
\caption{The ratio of $^{12}$C$(e,e'pN)$ double knockout events to $^{12}$C$(e,e'p)$ single knockout events, shown as a function 
of the reconstructed initial (missing) momentum of the knocked-out proton from the $^{12}$C$(e,e'p)$ reaction. Triangles 
and circles mark $^{12}$C$(e,e'pn)$and $^{12}$C$(e,e'pp)$ events, respectively. The square shows the $^{12}$C$(e,e'pp)/^{12}$C$(e,e'pn)$ ratio.  
A clear dominance of $^{12}$C$(e,e'pn)$ events is observed, evidence of the tensor nature of the nucleon-nucleon interaction 
in the measured momentum range. The pie chart on the right illustrates our understanding of the structure of $^{12}$C, 
composed of $80\%$ mean-field nucleons and $20\%$ SRC pairs, where the latter is composed of $\sim90\%$ $np$-SRC pairs 
and $~5\%$ $pp$ and $nn$ SRC pairs each.
(Figure reprinted with permission from American Association for Advancement in Science~\cite{Subedi:2008zz}.} 
\label{fig:HallA_Ratios}
\end{figure}

Two nucleon knockout experiments, measuring the $^{12}$C$(p,2pn)$ and
$^{12}$C$(e,e'pN)$ reactions, were done at BNL and JLab,
respectively~\cite{Tang:2003, Piasetzky:2006ai, Shneor:2007,
  Subedi:2008zz}. These experiments scattered protons and electrons
off high initial momentum ($300\le P_{initial} \le 600$ MeV/c) protons
in $^{12}$C and looked for the emission of a correlated recoil nucleon. 
In the absence of FSI, the initial momentum of the struck nucleon equals 
the missing momentum of the $^{12}C(e,e'p)$ and $^{12}C(p,2p)$ reactions. 
FSI will make this relationship more approximate. For simplicity, we will ignore FSI
The JLab measurement was sensitive to both proton and neutron
recoils but the BNL measurement was only sensitive to recoiling
neutrons. These experiments were performed at large momentum transfer
($Q^2\approx2$ GeV/c$^{2}$) where competing effects such a
Meson-Exchange-Currents (MEC) and Isobar Contributions (IC) are
suppressed and FSI are mainly confined to be between the nucleons of
the pair.  The main results of these experiments are shown in
Figs.~\ref{fig:OpeningAngle} and~\ref{fig:HallA_Ratios}.
Fig.~\ref{fig:OpeningAngle} shows the distribution of the cosine of
the opening angle between the initial momentum of the knocked-out
proton and the recoil nucleon.  The c.m. motion of the pairs in both
cases was found to be consistent with a gaussian in each direction,
with $\sigma=143\pm17$ (BNL) and $\sigma=136\pm20$ (JLab).  The BNL
results show a clear threshold around the fermi momentum where
recoiling neutrons above this momentum show a clear angular
correlation, and those below it do not. Fig.~\ref{fig:HallA_Ratios}
shows the ratio of single nucleon knockout events to two nucleon
knockout events, corrected for finite acceptance effects, as a
function of the initial momentum of the knocked-out proton. As can be
seen, within statistical uncertainties, all single proton knockout
events were accompanied by the emission of a recoil nucleon. The ratio
of proton recoil to neutron recoil was found to be approximately
1:20~\cite{Subedi:2008zz}. This is a clear evidence of the importance
of the tensor part of the nucleon-nucleon interaction at these
momentum scales~\cite{Sargsian:2005ru, Schiavilla:2006xx}. 

The effect of these measurements on our understanding of the short
distance nuclear structure is illustrated by the pie chart shown in
Fig.~\ref{fig:HallA_Ratios}. From the inclusive cross section ratio
measurements and from $A(e,e'p)$ measurements we know that in medium and heavy nuclei (i.e., $A\ge12$)
$\sim75-80\%$ of the nucleons are 'Mean-Field' nucleons, whereas
$\sim20-25\%$ have momentum greater than the Fermi momentum of the
nucleus.  Combined with results from exclusive two-nucleon knockout
measurements we know that these high momentum nucleons are dominated
by $2N$-SRC pairs, which are in turn dominated by neutron-proton
pairs.

\section{Short Range Correlations and the EMC Effect}
\label{sec:EMC-SRC}

\subsection{The EMC-SRC Correlation}
\label{sec:EMC-SRC_intro}


Analysis of world data on inclusive DIS and QE scattering cross section ratios showed that the magnitude of the EMC effect in nucleus $A$ is linearly related to the probability that a nucleon in that nucleus is part of a $2N$-SRC pair, see Fig.~\ref{fig:emcsrc} \cite{weinstein11,Hen12}. Here we used the $x_A$ corrected EMC data-base shown in~\ref{Appendix:EMC_DataBase} and defined the magnitude of the EMC effect, following Ref.~\cite{Seely:2009gt}, as the slope of the ratio of the per-nucleon DIS cross section of nucleus $A$ relative to deuterium, $dR_{\rm EMC}/dx$, in the region $0.35\le x_A \le 0.7$.  The probability that a nucleon belongs to an SRC pair is characterized by the SRC scale factor, $a_2(A/d)$, the ratio in the plateau region ($Q^2 \ge 1.5$ (GeV/$c$)$^2$ and $\xB\ge 1.5$) of the per-nucleon QE $(e,e')$ cross sections for nucleus $A$ and deuterium.

\begin{figure}[t]
  \centering
    \includegraphics[width=10cm, height=9cm]{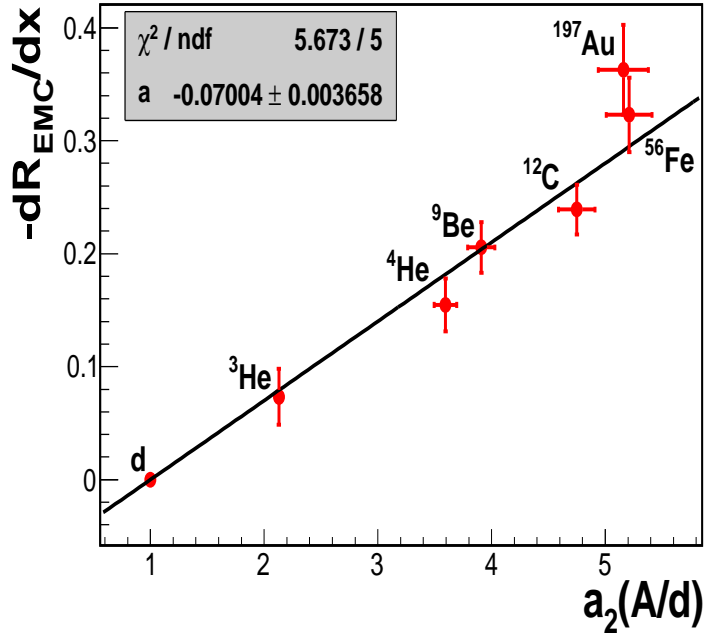} 
    \caption{
      The slope of the EMC effect 
      ($R_{\rm EMC}$, ratio of nuclear to deuteron cross section)  for $0.35 \le x_A \le 0.7$ plotted vs. $a_2(A/d)$, the SRC scale factor (the relative probability that a nucleon belongs to an SRC $NN$ pair) for a variety of nuclei \cite{Hen12}.  The fit parameter, $a=-0.084\pm0.004$ is the intercept of the line constrained to pass through the deuteron (and is therefore also the negative of the slope of that line).  }
     \label{fig:emcsrc} 
\end{figure}

The EMC effect correlates imperfectly with other $A$-dependent quantities (see~\cite{Seely:2009gt, Arrington12} and references therein).  In general, nuclei with $A\ge 4$ fall on one straight line but deuterium and $^3$He do not.  This is true when the EMC effect is plotted versus $A$, $A^{-1/3}$, or the average nuclear separation energy.  When plotting the EMC effect versus average nuclear density, $^9$Be is a clear outlier. This indicates that the excellent correlation with the SRC scale factor is not just a trivial byproduct of their mutual $A$ dependence.
 
The correlation between the EMC effect and the SRC scale factor is
robust \cite{Hen12}.  It applies to both new SRC data sets of Egiyan
\etal{}~\cite{Egiyan:2006}, and Fomin \etal{}~\cite{Fomin:2012}.  The
quality of the correlation also  does not depend on the corrections
applied to the SRC data.  These corrections include 
isoscalar cross section corrections, center-of-mass motion corrections and isoscalar pair-counting corrections.  The isoscalar correction to the SRC scale factors accounts for the different elementary electron-neutron and electron-proton cross sections.  This has a negligible effect on the fit quality and the extracted fit parameter.  Fomin \etal{} did not apply this correction, arguing that short range correlations are dominated by $np$ pairs.  Fomin \etal{}  argued that the SRC scale factors measured the relative probability of finding a high-momentum nucleon in nucleus $A$ relative to deuterium and that these scale factors needed to be corrected for the center-of-mass (cm) motion of the pair in order to determine the relative probability that a nucleon in nucleus $A$ belongs to an SRC pair.  As shown in both \cite{Hen12} and \cite{Arrington12}, including the pair c.m. motion correction improves the EMC-SRC correlation only slightly.  

This EMC-SRC correlation gives new insight into the origin of the EMC effect. Many different explanations of the EMC effect have been proposed since 1983.  After accounting for the standard nuclear effects of binding energy and fermi motion, explanations for the EMC effect fall into two general categories, those that require modifications of mean-field nucleons and those that require modifications of high-momentum nucleons.  The linear correlation between the strength of the EMC effect and the SRC scale factors indicates that possible modifications of nucleon structure occurs in nucleons belonging to SRC pairs.  This implies that the EMC effect, like short range correlations, is a short-distance, high virtuality, and high density phenomenon. 

Additionally, one can use the EMC-SRC correlation as a
phenomenological tool to constrain the deuteron IMC
effect\footnote{The deuteron In Medium Correction (IMC) effect was
  first introduced in Ref.~\cite{weinstein11} and refers to the
  difference between the DIS cross section for the deuteron and the sum of the cross sections for a free proton and neutron.}, and thus extract the free neutron structure function. Following Weinstein \etal{} \cite{weinstein11}, we can extrapolate the linear fit to the EMC-SRC correlation to the limit of $a_2(A/d)\rightarrow0$.  If the EMC effect and the SRC scale factor both stem from the same cause, then both will vanish at the same point.  The value $a_2(A/d)\rightarrow0$ is the limit of free nucleons with no SRC.  The extrapolation to the $y$-axis gives 
  $dR_{\rm EMC}/dx=-0.070\pm0.004$.  Since the EMC effect is linear for $0.3\le x_A \le0.7$ for all nuclei with $A>2$, we assume that the EMC effect is also linear in this region for the free proton plus neutron.  This gives the EMC effect for the free proton plus neutron:
\[
\frac{\sigma_d}{\sigma_p+\sigma_n} = 1 - a(\xB-b) \quad \hbox{for } 0.3\le\xB\le 0.7,
\]
where $\sigma_d$ and $\sigma_p$ are the measured DIS cross sections for the deuteron and free proton, $\sigma_n$ is the free neutron DIS cross section that we want to extract, $a = \vert dR_{\rm EMC}\vert =0.070\pm0.004$ and $b = 0.34\pm0.02$ is the average value of $\xB$ where the EMC ratio is unity\footnote{While the $x_A$ correction does not change much the slope of the EMC-SRC correlation, the $b$ parameter extracted here, while consistent within errors, is larger than that reported in Ref.~\cite{weinstein11}}. This implies that $\sigma_d/(\sigma_p+\sigma_n)$ decreases linearly from 1 to 0.97 as $x_p$ increases from 0.3 to 0.7.  We can then use this relationship to extract the free neutron cross section in this $x_p$ range. Incorporating this free neutron DIS cross section into the global QCD analysis~\cite{Accardi:2011fa}, one can better constrain the $d/u$ ratio at $x_p\rightarrow1$ to be equal to $0.23\pm0.09$ at the 90\% confidence level~\cite{hen11}.

The uncertainty quoted above is the uncertainty due to the data and the fit.  It does not include the uncertainty of corrections to the data.  As stated above, if we include  the correction for the cm motion of the correlated pair, then the fit parameter increases by 25\% and so does the free proton plus neutron EMC effect.  Arrington \etal{} claim that if we also consider including the isospin pair counting correction and alternative fitting methods, then the range of fits expands to $0.59 \le a \le 1.04$. The effect of these uncertainties on the extraction of the free neutron structure function and the $d/u$ ratio at large $x_p$ are discussed in Ref.~\cite{Piasetzky:2012yp}.

\subsection{Mean-Field versus Short Ranged Correlation contributions to the EMC Effect}
\label{sec:EMC-SRC_Theory}

We want to know whether the linear relation between the EMC slope and
the SRC plateau parameter $a_{2}(A/d)$ is more than a coincidence.
Any of the nuclear models discussed in Sec.~\ref{sec:SRC_Theory} has
correlations that would yield a value of $a_{2}(A/d)$ roughly
consistent with the measured values. None of these models incorporate
quark modifications of nuclear structure.  Therefore existence of $NN$
correlations is not a sufficient condition for the EMC effect
to occur.  The key questions are whether the quarks confined in
the two nucleons in an  SRC pair have different distribution functions
than those of two free nucleons. Thus the minimum input necessary to
test the existence of a relation between SRC and EMC is a model of a
modified two-nucleon structure function consistent with a good nuclear
model of SRC and with the EMC data.  Here we make a first attempt at
providing a link between SRC and EMC, by seeing if a modified
two-nucleon structure function associated with the
short-range-correlations can be used to describe deep inelastic
scattering on nuclei. We also consider the other possibility, that
medium modifications associated with the mean-field aspect of nuclei
can describe nuclear DIS.

The treatment of Frankfurt \& Strikman (FS)~\cite{Frankfurt:1985ui} is very useful for such an aim because  the nuclear structure information needed to compute deep inelastic scattering is encoded in only  three integrals that can be evaluated reliably.
 FS derive a convolution formula
\bea 
{1\over A}F_{2}^{A}(x_{A},Q^2)=\int_0^A \alpha\rho_A(\alpha)F_{2}^{N}(x_{A}/\alpha,Q^2)d\alpha,
\label{conv}
\eea
where $\alpha\equiv {A k\cdot q\over p_A\cdot q}$ is the fraction of the plus component of the nucleon momentum, with $k$ the struck nucleon initial momentum and $p_A=(m_A,0)$ is the nucleus 4-momentum.
 $ \rho_A(\alpha)$ is the probability that a nucleon in the nucleus
 carries momentum fraction $\alpha$
and $F_{2}^{N}$ is the free nucleon structure function ($F_{2}^{N}=\frac{1}{2}(F_{2}^{p}+F_{2}^{n})$).

Specifically $\rho(\alpha)$ is computed from the non-relativistic structure function,  $S_{A,NR}(k)$
\bea 
&&\rho_A(\alpha)=\int d^4kS_{A,NR}(k)\delta(\alpha-{k^0+k^3\over m_N}),
\eea
where 
\bea 
S_{A,{\rm NR}}(k,E)\equiv \langle A|a_k^\dagger \delta\left(E-H\right)a_k|A\rangle.
\label{spf}
\eea
The function $\rho(\alpha)$ is narrowly peaked about unity, so FS expand the nucleon structure function appearing in 
\eq{conv} about $\alpha=1$ to find:
\bea 
{1\over A}F_{2}^{A}(x_A)\approx F_{2}^{N}(x_A)I_1(A) +
x_{A} F'^{N}_{2}I_2(A)
+[x_{A} F'^{N}_{2}+{1\over2}x_{A}^2 F''^{N}_{2}]I_3(A)
,\label{conv1}
\eea
where for simplicity we neglect the $Q^2$ term in the structure function notation. The integrals $I_n(A)$ are given by 
\bea 
&&I_n(A)\equiv \int\rho_A(\alpha)\alpha (1-\alpha)^{n-1}d\alpha,\;n=1,2,3 
\eea
Note that  $I_1(A)=1$, which  is the normalization condition. 

FS proceed (see also~\cite{CiofidegliAtti:1991ae}) by isolating the
leading relativistic corrections of order $\epsilon_A/m$ and $k^2/m^2$ and
then using the Koltun sum rule~\cite{Koltun:1972kh} to find 
\bea
&&n_A(k)\equiv \langle A|a_k^\dagger a_k|A\rangle,\;I_1(A)=\int d^3kn_A(k).\\
&&I_2(A)=\int d^3k n_A(k)(2\epsilon_A/m+{A-4\over A-1}k^2/6m^2)\equiv{2\epsilon_A\over m}+{A-4\over A-1}\langle {k^2\over 6m^2}\rangle,\\
 &&I_3(A)=\int d^3k n_A(k)k^2/3m^2=\langle{k^2\over 3m^2}\rangle.
\eea
FS used the above formalism to show that a nucleons-only model without modified structure functions could not reproduce the deep inelastic scattering data.

To proceed with the calculation, we need a model of $n_A(k)$. The model of Ciofi degli Atti \& Simula~\cite{CiofidegliAtti:1995qe} is ideal for our purposes.
This is based on a realistic nuclear calculation of the spectral function that  leads to nuclear densities that yield qualitative  agreement with 
quasi elastic electron scattering. The model yields reasonably good agreement with the plateau values $a_2(A/d)$. Furthermore the
contributions of the mean-field and  correlation terms are enumerated in terms of the intermediate-state  energy $E$ appearing in the spectral function of \eq{spf}. 
This association with continuum energies, $E$, above about $20$ MeV with short range correlations is approximate but sufficiently accurate for the present schematic calculation. 
The spectral function leads to the momentum probability $n_A(k)$ such that
\bea 
n_A(k)=n_A^{(0)}(k)+ n_A^{(1)}(k),
\label{sep}
\eea
where the superscript 0 refers to that obtained from low energy terms dominated by the nuclear mean field and the superscript 1 refers to high energy terms (above the continuum threshold)  dominated by the effects of nucleon-nucleon correlations.  Ciofi degli Atti \& Simula provide functional forms for   $n_A(k)$ for several different nuclei. 
This separation using the excitation energy is not exactly the same as a separation in terms of relative momentum but is  qualitatively reasonable.
With this separation, terms involving correlations have about 20\% of the probability. 

 Using \eq{sep} one can obtain the separate contributions to $I_n(A)$ as $I_n(A)=I^{(0)}_n(A)+I^{(1)}_n(A).$

\begin{figure}[thbp!]
\centerline{\includegraphics[height=7.5cm,width=10cm]{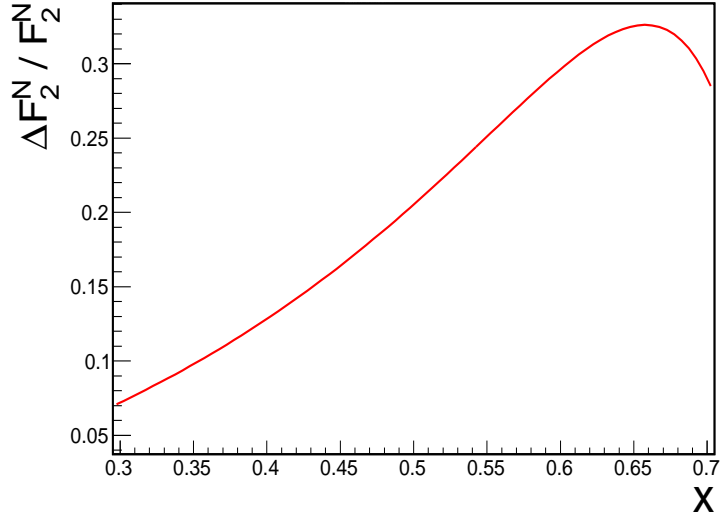}}
\caption{The ratio of the modification term, $\Delta F_{2}^{N}$ to the free nucleon structure function, $F_2^N$.}
\label{df_SRC}
\end{figure}

 \begin{figure}[thbp!]
\centerline{\includegraphics[width=13cm, height=20cm]{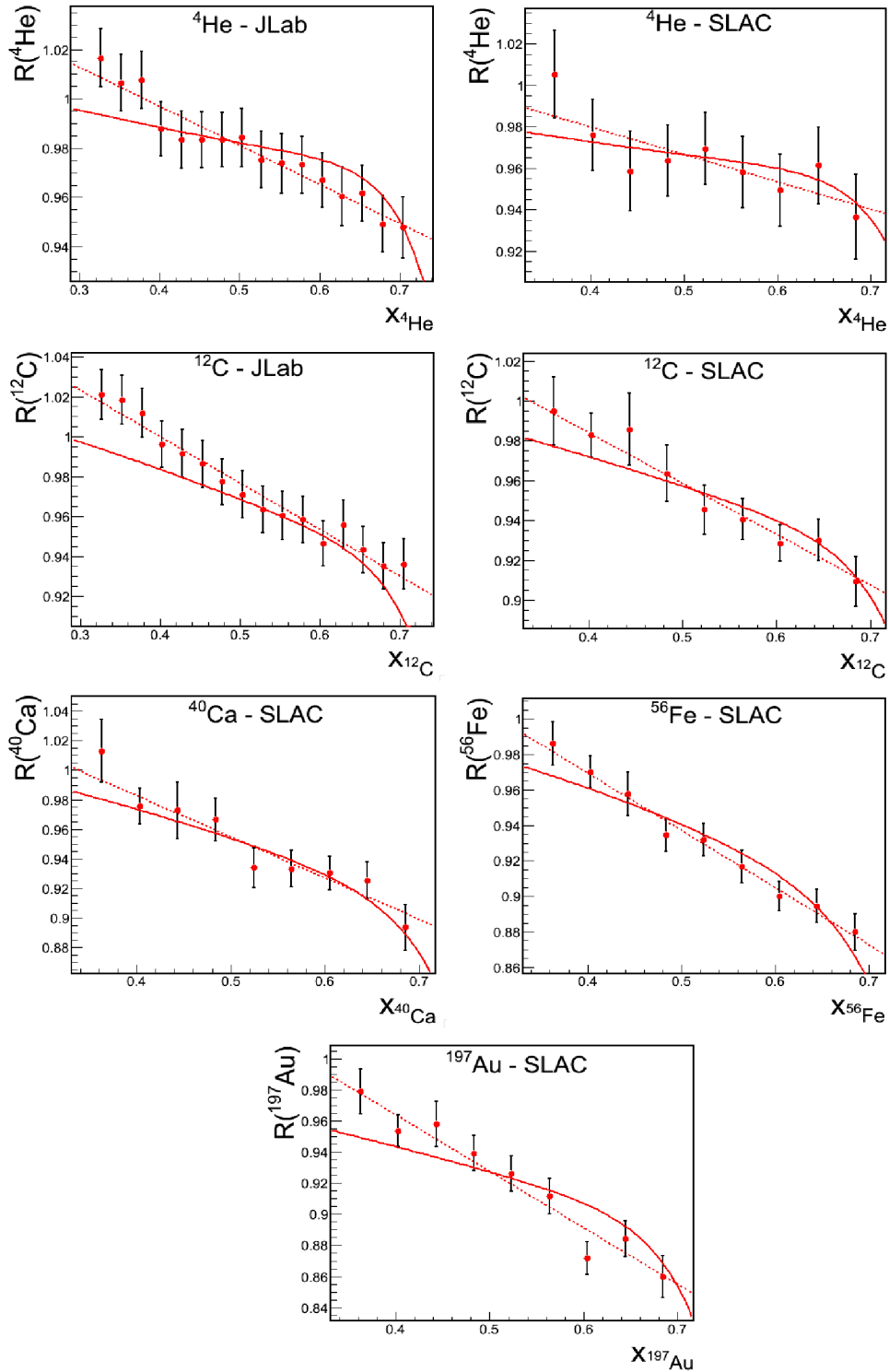}}
\caption{The ratios of free to bound structure functions for various nuclei, extracted in the nucleus reference frame as detailed in Eq.~\ref{eq:EMC_xACorrection}. The dashed line is the result of a linear fit to the data. The solid red line is the result of the medium-modification fit, assuming 
an $A$-independent modification to SRC nucleons.}
\label{fits_SRC}
\end{figure}

We next proceed by  assuming that nucleons in high energy excited states (correlated nucleons) have a different  structure function $\widetilde{F}_{2N}(x)$ than free ones $F_{2N}(x)$. Thus we make the replacements
\bea
I_1(A)F_2\rightarrow I^{(0)}_1(A)F_{2}^{N}+I^{(1)}_1(A)\widetilde{F}_{2}^{N}=I_1^{(0+1)}(A) F_{2}^{N}+I_1^{(1)}(A)\Delta F_{2N},\;{\rm etc},
\label{usesrc}
\eea
where
\bea 
\Delta F_{2}^{N}(x_A)=\widetilde{F}_{2}^{N}(x_A)-F_{2}^{N}(x_A).
\eea
An alternate version in which the medium modification is associated with the mean-field components of the  density can be obtained by using
\bea 
I_1(A)F_2^N\rightarrow I^{(0)}_1(A)\widetilde{F}_{2}^{N}+I^{(1)}_1(A)=I_1(A) F_{2}^{N}+I_1^{(0)}(A)\Delta F_{2}^{N},\;{\rm etc}.
\label{usemf}
\eea
A condition on $\Delta F_{2}^{N}$ derived from the baryon sum rule is that $\int_0^2dx_A{\Delta F_{2}^{N}(x_A)\over x_A}=0$. This means that 
$\Delta F_{2}^{N}$ must pass through 0 at some value of $x_A$.

The analysis proceeds by calculating Eq.~\ref{conv1} with the supplement of Eq.~\ref{usesrc} (Eq.~\ref{usemf} for the case of Mean-Field modification), assuming $\Delta F_{2}^{N}(x_A)$ is a second order polynomial in $x_A$. The parameters of $\Delta F_2^N(x_A)$ are fitted to the $x_A$ corrected EMC data (see~\ref{Appendix:EMC_DataBase}) for all nuclei for which momentum distributions are available (i.e., $^4$He, $^{12}$C, $^{40}$Ca, $^{56}$Fe, and $^{197}$Au). Note that the functional form of $\Delta F_2^N(x_A)$ is assumed to be 
independent of  $A$.

\begin{figure}[thbp!]
\centerline{\includegraphics[height=7.5cm,width=10cm]{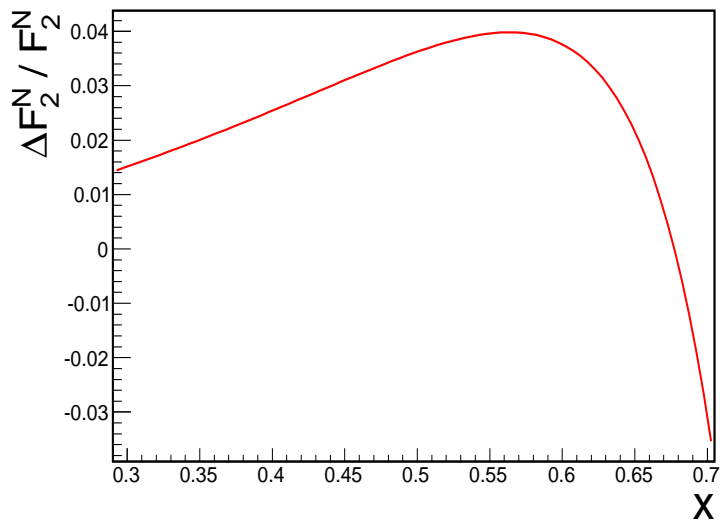}}
\caption{Same as Fig.~\ref{df_SRC}, assuming universal modification to Mean-Field nucleons. It is assumed that deuterium has no Mean-Field component, see text for details}
\label{df_MF}
\end{figure}

\begin{figure}[thbp!]
\centerline{\includegraphics[width=13cm, height=20cm]{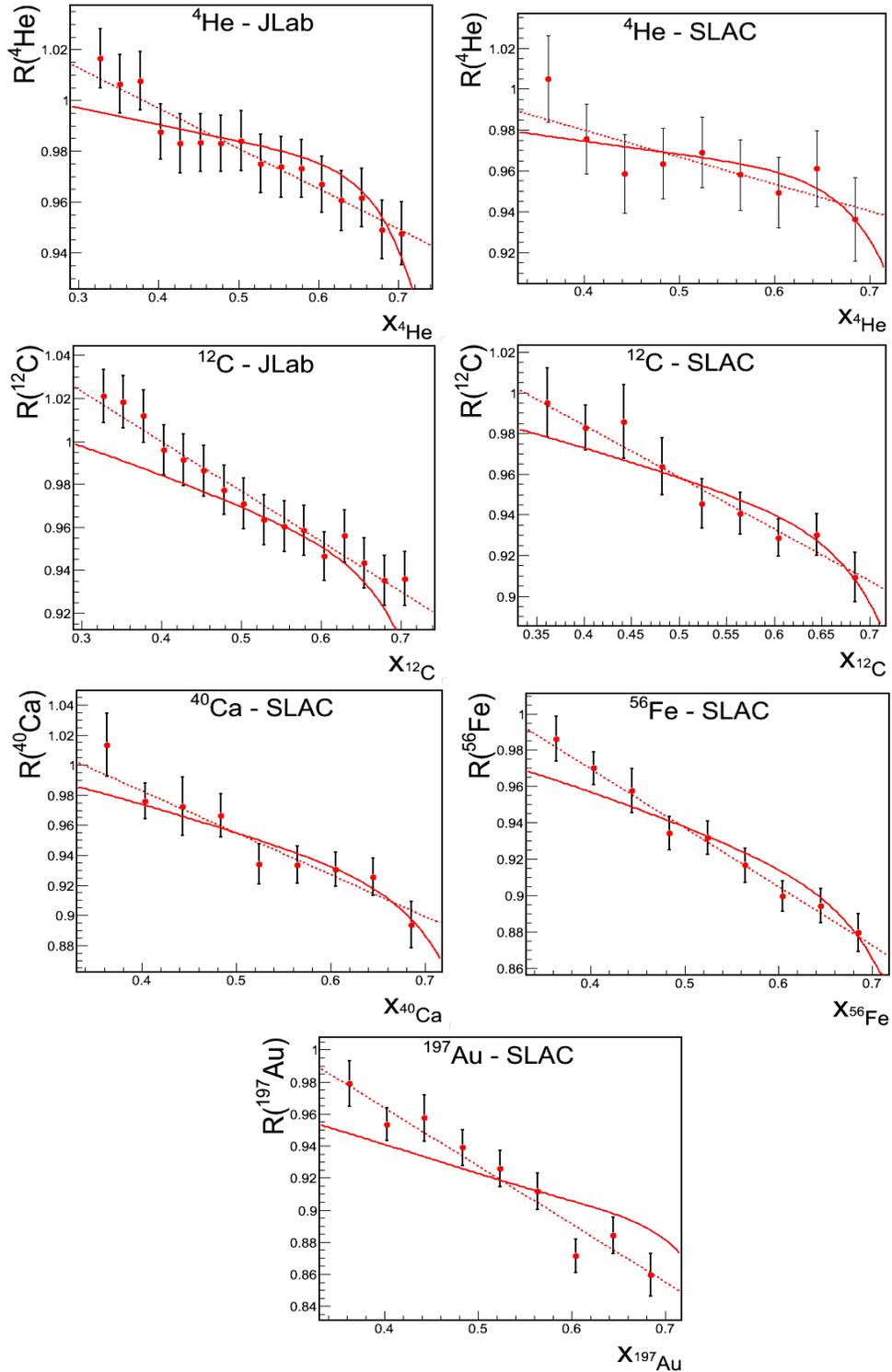}}
\caption{Same as Fig.~\ref{fits_SRC}, assuming universal modification to Mean-Field nucleons. It is assumed that deuterium has no Mean-Field component, see text for details.}
\label{fits_MF}\end{figure}

The results of the fits for individual nuclei are shown in Fig.~\ref{fits_SRC} (Fig.~\ref{fits_MF} for the case of Mean-Field modification). The description of the data is very good for all nuclei with a $\chi^2$ per degree of freedom of $\approx1$ for both the SRC and Mean-Field fits. These results were obtained using the parametrization of Ref.~\cite{Carlson:1983fs} for the free-nucleon structure function, $F_{2}^{N}$. The modified-to-free structure function ratio is shown in Fig.~\ref{df_SRC} (Fig.~\ref{df_MF} for the case of Mean-Field modification). 

The present results show that a model incorporating either universal
modification of Mean-Field nucleons or modification of nucleons in SRC
pairs can explain the EMC effect.  As expected, the required medium
modification of Mean-Field nucleons is on the order of a few percent
while that of SRC nucleons is a few tens of percent. This model does
not prove or disprove that the underlying cause of the EMC effect is
the unique association with short ranged correlations.  Note that
$^9$Be was not included in the model calculations since a $^9$Be
spectral function was not available. 
 Note also that this model does
not separate valence and sea quark distributions and therefore can't make predictions about the Drell-Yan data. 

Further experiments are needed to determine whether the Mean-Field or
SRC nucleons are modified by the nuclear medium. For example,
quasi-elastic electron scattering would be sensitive to the former but
not the latter.

\section{Summary}
We have reviewed recent data showing that the detailed $A$ dependence
of the EMC effect provides important hints in understanding the origin
of that 
effect. The EMC effect seems to depend on local density rather
than the average density.  We present the EMC ratio data (the ratio of nuclear to
deuterium structure functions) in terms of an improved Bjorken variable
$x_A=AQ^2/(2M_A q_0)$ (see the Appendix). We review 
short-range correlation data and discuss the linear relation
between the EMC effect and short-range correlations. We present a 
phenomenological model including modification of either Mean-Field or
SRC nucleons and find that modification of either is capable of accounting for the existing
data.
\label{sec:Summary}


\section*{Acknowledgements}
This work  is supported in part by the Israel Science Foundation, and the US-Israeli Bi-National Science Foundation, and the United States Department of Energy, Grants No. FG02-97ER41014 and DE-FG02-96ER40960.


\bibliographystyle{ws-ijmpe}
\bibliography{EMC_reviewMar15b}


\appendix
\section {EMC Data Base}
\label{Appendix:EMC_DataBase}
In this appendix we present a new data-base for the structure function ratio of nuclei ($F_2^A$) relative to deuterium ($F_2^d$). The ratio is taken using $F_2^A$ and $F_2^d$ each extracted in its own reference frame, at equivalent kinematical regions, defined by $x_A$ and $Q^2$. The extraction of the structure function follows the formalism of section~\ref{sec:EMC_Data}, and is shown only for the range of $0.3\le x_A \le 0.7$. 

We use JLab and SLAC high precision data on DIS cross section ratios for nuclei relative to deuterium as input to Eq.~\ref{eq:EMC_xACorrection}~\cite{Gomez:1993ri,Seely:2009gt}. We use a parametrization of $F_2^d(x,Q^2)$  from ref~\cite{Bosted:2008} to move the deuteron measurement to $x_A$, and reapply isoscalar corrections using a parametrization of $R_{np}(x_A,Q^2)={F_2^n(x_A,Q^2)}/{F_2^p(x_A,Q^2)}$ from ref~\cite{Arrington:2009}. The results are presented in tables~\ref{tab:4He_EMC_SLAC}-~\ref{tab:197Au_EMC_SLAC} are for the SLAC data and tables~\ref{tab:3He_EMC_JLab}-~\ref{tab:12C_EMC_JLab} are for the JLab data.
Note that the SLAC results are averaged over $Q^2=2$, $5$, and $10$ GeV/c$^{2}$.


\begin{table}[h]
\caption{EMC data for $^4$He from SLAC. The left side of the table shows the original as published data from table VIII of ref~\cite{Gomez:1993ri}. The right side is the same data, corrected for the definition of $x_A$ according to Eq.~\ref{eq:EMC_xACorrection}.}
\raggedright
\begin{tabular}{|c|c|c || c|c|c|}
\hline
           x       &     $F_2^{^4He}/F_2^d(x)$     &    $\Delta F_2^{^4He}/F_2^d(x)$    &      $x_A$        &   $F_2^{^4He}/F_2^d(x_A)$  &      $\Delta F_2^{^4He}/F_2^d(x_A)$       \\
     \hline
	0.36	 & 		0.998			 & 		0.021				 & 	0.362	 & 		1.005			 & 	0.021 \\
	0.40	 & 		0.968			 & 		0.017				 & 	0.402	 & 		0.976			 & 	0.017 \\ 
	0.44	 & 		0.949			 & 		0.019				 & 	0.443	 & 		0.958			 & 	0.019 \\
	0.48	 & 		0.954			 & 		0.017				 & 	0.483	 & 		0.963			 & 	0.017 \\
	0.52	 & 		0.951			 & 		0.017				 & 	0.523	 & 		0.969			 & 	0.017 \\
	0.56	 & 		0.943			 & 		0.017				 & 	0.563	 & 		0.958			 & 	0.017 \\
	0.60	 & 		0.928			 & 		0.017				 & 	0.604	 & 		0.949			 & 	0.017 \\
	0.64	 & 		0.935			 & 		0.018				 & 	0.644	 & 		0.961			 & 	0.018 \\
	0.68	 & 		0.917			 & 		0.020				 & 	0.684	 & 		0.936			 & 	0.020 \\
    \hline
\end{tabular}
\label{tab:4He_EMC_SLAC}
\end{table}

\begin{table}[h]
\caption{Same as table~\ref{tab:4He_EMC_SLAC}, for $^{9}$Be.}
\raggedright
\begin{tabular}{|c|c|c || c|c|c|}
\hline
           x       &     $F_2^{^9Be}/F_2^d(x)$     &    $\Delta F_2^{^9Be}/F_2^d(x)$    &      $x_A$        &   $F_2^{^9Be}/F_2^d(x_A)$  &      $\Delta F_2^{^9Be}/F_2^d(x_A)$       \\
     \hline
	0.36	 & 		0.993			 & 		0.014				 & 	0.362	 & 		1.007	 		& 	0.014 \\
	0.40	 & 		0.957			 & 		0.009				 & 	0.402	 & 		0.972	 		& 	0.009 \\
	0.44	 & 		0.980 			& 		0.014	 			& 	0.443	 & 		0.997	 		& 	0.014 \\
	0.48	 & 		0.951			 & 		0.011				 & 	0.483	 & 		0.968	 		& 	0.011 \\
	0.52	 & 		0.955			 & 		0.011				 & 	0.523	 & 		0.979	 		& 	0.011 \\
	0.56	 & 		0.945			 & 		0.011				 & 	0.563	 & 		0.967	 		& 	0.011 \\
	0.60	 & 		0.928			 & 		0.010	 			& 	0.604	 & 		0.955	 		& 	0.010 \\
	0.64	 & 		0.917			 & 		0.011				 & 	0.644	 & 		0.947	 		& 	0.011 \\
	0.68	 & 		0.912			 & 		0.013				 & 	0.684	 & 		0.935	 		& 	0.013 \\
    \hline
\end{tabular}
\label{tab:9Be_EMC_SLAC}
\end{table}

\begin{table}[h]
\caption{Same as table~\ref{tab:4He_EMC_SLAC}, for $^{12}$C.}
\raggedright
\begin{tabular}{|c|c|c || c|c|c|}
\hline
           x       &     $F_2^{^{12}C}/F_2^d(x)$     &    $\Delta F_2^{^{12}C}/F_2^d(x)$    &      $x_A$        &   $F_2^{^{12}C}/F_2^d(x_A)$  &      $\Delta F_2^{^{12}C}/F_2^d(x_A)$       \\
     \hline
	0.36	 & 		0.987			 & 		0.017				 & 	0.362		 & 		0.995			 & 	0.017 \\
	0.40	 & 		0.974			 & 		0.011				 & 	0.403		 & 		0.983			 & 	0.011 \\
	0.44	 & 		0.975			 & 		0.018				 & 	0.443		 & 		0.986			 & 	0.018 \\
	0.48	 & 		0.953			 & 		0.014				 & 	0.483		 & 		0.963			 & 	0.014 \\ 
	0.52	 & 		0.926			 & 		0.012				 & 	0.523		 & 		0.945			 & 	0.012 \\
	0.56	 & 		0.924			 & 		0.010				 & 	0.564		 & 		0.940			 & 	0.010 \\
	0.60	 & 		0.905			 & 		0.009				 & 	0.604		 & 		0.928			 & 	0.009 \\
	0.64	 & 		0.903			 & 		0.010				 & 	0.644		 & 		0.930			 & 	0.010 \\
	0.68	 & 		0.888			 & 		0.012				 & 	0.685		 & 		0.909			 & 	0.012 \\
    \hline
\end{tabular}
\label{tab:12C_EMC_SLAC}
\end{table}

\begin{table}[h]
\caption{Same as table~\ref{tab:4He_EMC_SLAC}, for $^{27}$Al.}
\raggedright
\begin{tabular}{|c|c|c || c|c|c|}
\hline
           x       &     $F_2^{^{27}Al}/F_2^d(x)$     &    $\Delta F_2^{^{27}Al}/F_2^d(x)$    &      $x_A$        &   $F_2^{^{27}Al}/F_2^d(x_A)$  &      $\Delta F_2^{^{27}Al}/F_2^d(x_A)$       \\
     \hline
	0.36	 & 		0.993			 & 		0.013				 & 	0.362		 & 		1.005			 & 	0.013 \\
	0.40	 & 		0.966			 & 		0.009				 & 	0.403		 & 		0.977			 & 	0.009 \\
	0.44	 & 		0.959			 & 		0.012				 & 	0.443		 & 		0.973			 & 	0.012 \\
	0.48	 & 		0.934			 & 		0.010				 & 	0.483		 & 		0.948			 & 	0.010 \\ 
	0.52	 & 		0.926			 & 		0.010				 & 	0.524		 & 		0.950			 & 	0.010 \\
	0.56	 & 		0.923			 & 		0.009				 & 	0.564		 & 		0.944			 & 	0.009 \\
	0.60	 & 		0.906			 & 		0.009				 & 	0.604		 & 		0.934			 & 	0.009 \\
	0.64	 & 		0.892			 & 		0.009				 & 	0.645		 & 		0.923			 & 	0.009 \\
	0.68	 & 		0.876			 & 		0.011				 & 	0.685		 & 		0.900			 & 	0.011 \\
    \hline
\end{tabular}
\label{tab:27Al_EMC_SLAC}
\end{table}

\begin{table}[h]
\caption{Same as table~\ref{tab:4He_EMC_SLAC}, for $^{40}$Ca.}
\raggedright
\begin{tabular}{|c|c|c || c|c|c|}
\hline
           x       &     $F_2^{^{40}Ca}/F_2^d(x)$     &    $\Delta F_2^{^{40}Ca}/F_2^d(x)$    &      $x_A$        &   $F_2^{^{40}Ca}/F_2^d(x_A)$  &      $\Delta F_2^{^{40}Ca}/F_2^d(x_A)$       \\
     \hline
	0.36	 & 		1.004			 & 		0.021				 & 	0.363		 & 		1.013				 & 	0.021 \\
	0.40	 & 		0.966			 & 		0.012				 & 	0.403		 & 		0.975				 & 	0.012 \\
	0.44	 & 		0.960			 & 		0.019				 & 	0.443		 & 		0.972				 & 	0.019 \\
	0.48	 & 		0.954			 & 		0.014				 & 	0.484		 & 		0.966				 & 	0.014 \\
	0.52	 & 		0.912			 & 		0.013				 & 	0.524		 & 		0.934				 & 	0.013 \\
	0.56	 & 		0.915			 & 		0.012				 & 	0.564		 & 		0.933				 & 	0.012 \\
	0.60	 & 		0.904			 & 		0.011				 & 	0.605		 & 		0.930				 & 	0.011 \\
	0.64	 & 		0.895			 & 		0.012				 & 	0.645		 & 		0.925				 & 	0.012 \\
	0.68	 & 		0.870			 & 		0.015				 & 	0.685		 & 		0.893				 & 	0.015 \\
    \hline
\end{tabular}
\label{tab:40Ca_EMC_SLAC}
\end{table}

\begin{table}[h]
\caption{Same as table~\ref{tab:4He_EMC_SLAC}, for $^{56}$Fe.}
\raggedright
\begin{tabular}{|c|c|c || c|c|c|}
\hline
           x       &     $F_2^{^{56}Fe}/F_2^d(x)$     &    $\Delta F_2^{^{56}Fe}/F_2^d(x)$    &      $x_A$        &   $F_2^{^{56}Fe}/F_2^d(x_A)$  &      $\Delta F_2^{^{56}Fe}/F_2^d(x_A)$       \\
     \hline
	0.36	 & 		0.972			 & 		0.012					 & 	0.363	 & 		0.986				 & 	0.012 \\
	0.40	 & 		0.955			 & 		0.009					 & 	0.403	 & 		0.970				 & 	0.009 \\
	0.44	 & 		0.940			 & 		0.012					 & 	0.443	 & 		0.957				 & 	0.012 \\
	0.48	 & 		0.917			 & 		0.009					 & 	0.484	 & 		0.934				 & 	0.009 \\
	0.52	 & 		0.904			 & 		0.009					 & 	0.524	 & 		0.931				 & 	0.009 \\
	0.56	 & 		0.893			 & 		0.009					 & 	0.564	 & 		0.916				 & 	0.009 \\
	0.60	 & 		0.869			 & 		0.008					 & 	0.605	 & 		0.899				 & 	0.008 \\
	0.64	 & 		0.860			 & 		0.009					 & 	0.645	 & 		0.894				 & 	0.009 \\
	0.68	 & 		0.852			 & 		0.010					 & 	0.685	 & 		0.879				 & 	0.010 \\
    \hline
\end{tabular}
\label{tab:56Fe_EMC_SLAC}
\end{table}

\begin{table}[h]
\caption{Same as table~\ref{tab:4He_EMC_SLAC}, for $^{108}$Ag.}
\raggedright
\begin{tabular}{|c|c|c || c|c|c|}
\hline
           x       &     $F_2^{^{108}Ag}/F_2^d(x)$     &    $\Delta F_2^{^{108}Ag}/F_2^d(x)$    &      $x_A$        &   $F_2^{^{108}Ag}/F_2^d(x_A)$  &      $\Delta F_2^{^{108}Ag}/F_2^d(x_A)$       \\
     \hline
	0.36	 & 		1.012			 & 		0.023					 & 	0.363		 & 		1.031			 & 	0.023 \\
	0.40	 & 		0.968			 & 		0.013					 & 	0.403		 & 		0.988			 & 	0.013 \\
	0.44	 & 		0.957			 & 		0.021					 & 	0.443		 & 		0.979			 & 	0.021 \\
	0.48	 & 		0.926			 & 		0.015					 & 	0.484		 & 		0.948			 & 	0.015 \\
	0.52	 & 		0.897			 & 		0.014					 & 	0.524		 & 		0.928			 & 	0.014 \\
	0.56	 & 		0.891			 & 		0.013					 & 	0.564		 & 		0.918			 & 	0.013 \\
	0.60	 & 		0.881			 & 		0.012					 & 	0.605		 & 		0.915			 & 	0.012 \\
	0.64	 & 		0.842			 & 		0.013					 & 	0.645		 & 		0.878			 & 	0.013 \\
	0.68	 & 		0.842			 & 		0.016					 & 	0.685	 	& 		0.871			 & 	0.016 \\
    \hline
\end{tabular}
\label{tab:108Ag_EMC_SLAC}
\end{table}

\begin{table}[h]
\caption{Same as table~\ref{tab:4He_EMC_SLAC}, for $^{197}$Au.}
\raggedright
\begin{tabular}{|c|c|c || c|c|c|}
\hline
           x       &     $F_2^{^{197}Au}/F_2^d(x)$     &    $\Delta F_2^{^{197}Au}/F_2^d(x)$    &      $x_A$        &   $F_2^{^{197}Au}/F_2^d(x_A)$  &      $\Delta F_2^{^{197}Au}/F_2^d(x_A)$       \\
     \hline
	0.36	 & 		0.956			 & 		0.014					 & 	0.362		 & 		0.979			 & 	0.014 \\
	0.40	 & 		0.930			 & 		0.010					 & 	0.403		 & 		0.953			 & 	0.010 \\
	0.44	 & 		0.931			 & 		0.014					 & 	0.443		 & 		0.957			 & 	0.014 \\
	0.48	 & 		0.914			 & 		0.011					 & 	0.483		 & 		0.939			 & 	0.011 \\
	0.52	 & 		0.892			 & 		0.011					 & 	0.524		 & 		0.926			 & 	0.011 \\
	0.56	 & 		0.881			 & 		0.011					 & 	0.564		 & 		0.911			 & 	0.011 \\
	0.60	 & 		0.837			 & 		0.010					 & 	0.604		 & 		0.871			 & 	0.010 \\
	0.64	 & 		0.846			 & 		0.011					 & 	0.644		 & 		0.884			 & 	0.011 \\
	0.68	 & 		0.829			 & 		0.013					 & 	0.685		 & 		0.859			 & 	0.013 \\
    \hline
\end{tabular}
\label{tab:197Au_EMC_SLAC}
\end{table}

\begin{table}[h]
\caption{EMC data for $^3He$ from JLab. The left side of the table shows the original as published data from ref~\cite{Seely:2009gt}. The right side is the same data, corrected for the definition of $x_A$ according to Eq.~\ref{eq:EMC_xACorrection}.}
\raggedright
\begin{tabular}{|c|c|c || c|c|c|}
\hline
           x       &     $F_2^{^3He}/F_2^d(x)$     &    $\Delta F_2^{^3He}/F_2^d(x)$    &      $x_A$        &   $F_2^{^3He}/F_2^d(x_A)$  &      $\Delta F_2^{^3He}/F_2^d(x_A)$       \\
     \hline
	0.325	 & 		0.9774			 & 		0.011453			         	 & 	0.325	 & 		0.970073			 & 	0.0113672 \\
	0.350	 & 		0.9763			 & 		0.0113158			 & 	0.350	 & 		0.968431			 & 	0.0112246 \\
	0.375	 & 		0.9796			 & 		0.0113219			 & 	0.375	 & 		0.971228			 & 	0.0112251 \\ 
	0.400	 & 		0.9684			 & 		0.0107239			 & 	0.400	 & 		0.959746			 & 	0.0106281 \\
	0.425	 & 		0.9725			 & 		0.0114144			 & 	0.425	 & 		0.963504			 & 	0.0113089 \\
	0.450	 & 		0.9713			 & 		0.0112523			 & 	0.450	 & 		0.961924			 & 	0.0111437 \\
	0.475	 & 		0.9696			 & 		0.0108533			 & 	0.476	 & 		0.960588			 & 	0.0107524 \\
	0.500	 & 		0.9629			 & 		0.0114935			 & 	0.501	 & 		0.95354			 & 	0.0113818 \\
	0.525	 & 		0.9599			 & 		0.0112036			 & 	0.526	 & 		0.949848			 & 	0.0110863 \\
	0.550	 & 		0.964			 & 		0.0118444			 & 	0.551	 & 		0.955055			 & 	0.0117345 \\
	0.575	 & 		0.9653			 & 		0.0113391			 & 	0.576	 & 		0.955923			 & 	0.0112289 \\
	0.600	 & 		0.9644			 & 		0.0109817			 & 	0.601	 & 		0.954435			 & 	0.0108683 \\
	0.625	 & 		0.949			 & 		0.0118153			 & 	0.626	 & 		0.94051			 & 	0.0117096 \\
	0.650	 & 		0.9611			 & 		0.0115051			 & 	0.651	 & 		0.951985			 & 	0.0113959 \\
	0.675	 & 		0.9562			 & 		0.0116562			 & 	0.676	 & 		0.945647			 & 	0.0115276 \\
	0.700	 & 		0.9479			 & 		0.0125035			 & 	0.701	 & 		0.938561			 & 	0.0123803 \\
    \hline
\end{tabular}
\label{tab:3He_EMC_JLab}
\end{table}

\begin{table}[h]
\caption{Same as table~\ref{tab:3He_EMC_JLab}, for $^{4}$He.}
\raggedright
\begin{tabular}{|c|c|c || c|c|c|}
\hline
           x       &     $F_2^{^{4}He}/F_2^d(x)$     &    $\Delta F_2^{^{4}He}/F_2^d(x)$    &      $x_A$        &   $F_2^{^{4}He}/F_2^d(x_A)$  &      $\Delta F_2^{^{4}He}/F_2^d(x_A)$       \\
     \hline
	0.325	 & 		1.011			 & 		0.0116698				 & 	0.327	 & 		1.01653			 & 	0.0117336 \\
	0.350	 & 		0.9998			 & 		0.0114398				 & 	0.352	 & 		1.00644			 & 	0.0115158 \\
	0.375	 & 		0.9996			 & 		0.0114471				 & 	0.377	 & 		1.00753			 & 	0.0115379 \\
	0.400	 & 		0.9784			 & 		0.0108174				 & 	0.402	 & 		0.987558			 & 	0.0109187 \\
	0.425	 & 		0.9727			 & 		0.0114294				 & 	0.427	 & 		0.983079			 & 	0.0115513 \\
	0.450	 & 		0.9724			 & 		0.0112776				 & 	0.453	 & 		0.98331			 & 	0.0114042 \\
	0.475	 & 		0.9688			 & 		0.0108597				 & 	0.478	 & 		0.983158			 & 	0.0110206 \\
	0.500	 & 		0.9695			 & 		0.0115953				 & 	0.503	 & 		0.984013			 & 	0.0117688 \\
	0.525	 & 		0.9613			 & 		0.0112505				 & 	0.528	 & 		0.975085			 & 	0.0114118 \\
	0.550	 & 		0.955			 & 		0.0117935				 & 	0.553	 & 		0.973823			 & 	0.0120259 \\
	0.575	 & 		0.9542			 & 		0.0112231				 & 	0.578	 & 		0.973198			 & 	0.0114466 \\
	0.600	 & 		0.9491			 & 		0.0107922				 & 	0.604	 & 		0.966896			 & 	0.0109946 \\
	0.625	 & 		0.9361			 & 		0.0115938				 & 	0.629	 & 		0.960513			 & 	0.0118962 \\
	0.650	 & 		0.9389			 & 		0.0111987				 & 	0.654	 & 		0.961687			 & 	0.0114705 \\
	0.675	 & 		0.9315			 & 		0.0113139				 & 	0.679	 & 		0.949143			 & 	0.0115282 \\
	0.700	 & 		0.9238			 & 		0.0121262				 & 	0.704	 & 		0.947597			 & 	0.0124385 \\
    \hline
\end{tabular}
\label{tab:4He_EMC_JLab}
\end{table}

\begin{table}[h]
\caption{Same as table~\ref{tab:3He_EMC_JLab}, for $^{9}$Be.}
\raggedright
\begin{tabular}{|c|c|c || c|c|c|}
\hline
           x       &     $F_2^{^{9}Be}/F_2^d(x)$     &    $\Delta F_2^{^{9}Be}/F_2^d(x)$    &      $x_A$        &   $F_2^{^{9}Be}/F_2^d(x_A)$  &      $\Delta F_2^{^{9}Be}/F_2^d(x_A)$       \\
     \hline
	0.325	 & 		1.027			 & 		0.0134537				 & 	0.327	 & 		1.03628			 & 	0.0135753 \\
	0.350	 & 		1.018			 & 		0.0131816				 & 	0.352	 & 		1.02794			 & 	0.0133104 \\
	0.375	 & 		1.014			 & 		0.0130497				 & 	0.377	 & 		1.02575			 & 	0.0132009 \\
	0.400	 & 		0.9977			 & 		0.0124262				 & 	0.402	 & 		1.0112			 & 	0.0125944 \\
	0.425	 & 		0.9907			 & 		0.0127738				 & 	0.427	 & 		1.00495			 & 	0.0129575 \\
	0.450	 & 		0.9821			 & 		0.0124847				 & 	0.452	 & 		0.997484			 & 	0.0126803 \\
	0.475	 & 		0.9745			 & 		0.0120235				 & 	0.477	 & 		0.992554			 & 	0.0122463 \\
	0.500	 & 		0.9709			 & 		0.012429					 & 	0.503	 & 		0.989694			 & 	0.0126696 \\
	0.525	 & 		0.9567			 & 		0.0119992				 & 	0.528	 & 		0.974422			 & 	0.0122215 \\
	0.550	 & 		0.9538			 & 		0.0123411				 & 	0.553	 & 		0.976809			 & 	0.0126388 \\
	0.575	 & 		0.9469			 & 		0.0118303				 & 	0.578	 & 		0.969655			 & 	0.0121146 \\
	0.600	 & 		0.9403			 & 		0.0114402				 & 	0.603	 & 		0.961711			 & 	0.0117006 \\
	0.625	 & 		0.9459			 & 		0.0122301				 & 	0.628	 & 		0.974281			 & 	0.0125971 \\
	0.650	 & 		0.9322			 & 		0.0116412				 & 	0.653	 & 		0.958901			 & 	0.0119747 \\
	0.675	 & 		0.9269			 & 		0.0116589				 & 	0.679	 & 		0.948751			 & 	0.0119338 \\
	0.700	 & 		0.9201			 & 		0.0122438				 & 	0.704	 & 		0.947675			 & 	0.0126108 \\
    \hline
\end{tabular}
\label{tab:9Be_EMC_JLab}
\end{table}

\begin{table}[h]
\caption{Same as table~\ref{tab:3He_EMC_JLab}, for $^{12}$C.}
\raggedright
\begin{tabular}{|c|c|c || c|c|c|}
\hline
           x       &     $F_2^{^{12}C}/F_2^d(x)$     &    $\Delta F_2^{^{12}C}/F_2^d(x)$    &      $x_A$        &   $F_2^{^{12}C}/F_2^d(x_A)$  &      $\Delta F_2^{^{12}C}/F_2^d(x_A)$       \\
     \hline
	0.325	 & 		1.015		 & 		0.0123864				 & 	0.327	 & 		1.0211			 & 	0.0124609 \\
	0.350	 & 		1.011		 & 		0.0122230				 & 	0.352	 & 		1.01838			 & 	0.0123122 \\
	0.375	 & 		1.003		 & 		0.0120969				 & 	0.377	 & 		1.01175			 & 	0.0122024 \\
	0.400	 & 		0.9859		 & 		0.0115927				 & 	0.403	 & 		0.996045			 & 	0.011712 \\
	0.425	 & 		0.9798		 & 		0.0119255				 & 	0.428	 & 		0.991303			 & 	0.0120655 \\
	0.450	 & 		0.9743		 & 		0.0117119				 & 	0.453	 & 		0.986333			 & 	0.0118566 \\
	0.475	 & 		0.9617		 & 		0.0113265				 & 	0.478	 & 		0.977356			 & 	0.0115109 \\
	0.500	 & 		0.9553		 & 		0.0117368				 & 	0.503	 & 		0.971005			 & 	0.0119297 \\
	0.525	 & 		0.9485		 & 		0.0114379				 & 	0.528	 & 		0.963524			 & 	0.0116191 \\
	0.550	 & 		0.9401		 & 		0.0117128				 & 	0.554	 & 		0.960463			 & 	0.0119665 \\
	0.575	 & 		0.938		 & 		0.0113657				 & 	0.579	 & 		0.958495			 & 	0.0116141 \\
	0.600	 & 		0.9274		 & 		0.0110039				 & 	0.604	 & 		0.946495			 & 	0.0112305 \\
	0.625	 & 		0.9291		 & 		0.0119147				 & 	0.629	 & 		0.955914			 & 	0.0122586 \\
	0.650	 & 		0.9191		 & 		0.0113583				 & 	0.654	 & 		0.943392			 & 	0.0116585 \\
	0.675	 & 		0.9162		 & 		0.0114603				 & 	0.680	 & 		0.93512			 & 	0.011697 \\
	0.700	 & 		0.9107		 & 		0.0121902				 & 	0.705	 & 		0.936063			 & 	0.0125297 \\
    \hline
\end{tabular}
\label{tab:12C_EMC_JLab}
\end{table}

\end{document}